\documentclass{article}
\usepackage[version=3]{mhchem} 
\usepackage{siunitx}
\usepackage{subfig}
\usepackage{longtable}
\usepackage{stmaryrd}
\usepackage{placeins}
\usepackage{graphicx}

\usepackage{geometry}
\begin{document}
\begin{center}

\huge\textbf{Triplet Rydberg states of \\ aluminum monofluoride}\\
\vspace{3cm}

\large{N. Walter,$^{a,b}$,
M. Doppelbauer,$^a$
S. Schaller,$^a$
X. Liu,$^a$
R. Thomas,$^a$
S. Wright,$^a$
B.G. Sartakov,$^a$
G. Meijer,$^{a,c}$}

\vspace{2cm}

\large
{$^{a}$ Fritz Haber Institute of the Max Planck Society, Faradayweg 4--6, 14195 Berlin, Germany\\
$^{b}$ walter@fhi-berlin.mpg.de,   $^{c}$ meijer@fhi-berlin.mpg.de

}

\end{center}

\vspace{2cm}

Aluminum monofluoride (AlF) is a suitable molecule for laser cooling and trapping.
Such experiments require an extensive spectroscopic characterization of the electronic structure.
Two of the theoretically predicted higher lying triplet states of AlF, the counterparts of the well-characterized D$^1\Delta$ and E$^1\Pi$ states, had experimentally not been identified yet. We here report on the characterization of the d$^3\Pi$ ($v=0-6$) and e$^3\Delta$ ($v=0-2$) states, confirming the predicted energetic ordering of these states (J. Chem. Phys. 88 (1988) 5715-5725), as well as of the f$^3\Sigma^+$ ($v=0-2$) state. The transition intensity of the d$^3\Pi, v=3$ -- a$^3\Pi, v=3$ band is negligibly small. This band gets its weak, unexpected rotational structure via intensity borrowing from the nearby e$^3\Delta, v=2$ -- a$^3\Pi, v=3$ band, made possible via spin-orbit and spin-rotation interaction between the d$^3\Pi$ and e$^3\Delta$ states. This interaction affects the equilibrium rotational constants in both states; their deperturbed values yield equilibrium internuclear distances that are consistent with the observations. 
We determine the ionization potential of AlF to be 78492(1)\,cm$^{-1}$ by ionization from the d$^3\Pi$ state.

\begin{figure}[]
        \centering
        \includegraphics[width=350pt]{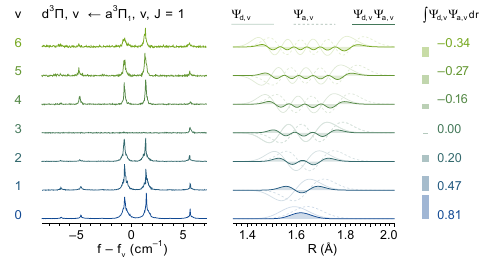}
        \label{fig:GA}
        \end{figure}
        

\FloatBarrier
\section{Introduction}
A first comprehensive overview of the electronic spectrum of gaseous AlF was given by Barrow, Kopp and Malmberg in 1974.\cite{Barrow1974a} At that time, a total of nine singlet states (including the X$^1\Sigma^+$ ground state) and seven triplet states had been assigned to AlF. These states were ordered into a common energy scheme and the nomenclature for some of the triplet states was revised to get a labeling in alphabetic order with increasing energy. A state that had been tentatively assigned as a $^3\Delta$ state\cite{So1974}, the counterpart of the well-characterized D$^1\Delta$ state, was designated as the d$^3\Delta$ state, but it was given a question mark. The rotational structure of this state had not been resolved and the state (originally denoted as c$^3\Sigma$) was only characterized by the energy of its lowest two vibrational levels.\cite{Dodsworth1955} In a theoretical study on the lowest six singlet and lowest five triplet states of AlF, the assignment of this state as a $^3\Delta$ state was substantiated, but it was predicted that there should be a $^3\Pi$ state, the counterpart of the E$^1\Pi$ state, about 2000\,cm$^{-1}$ lower in energy.\cite{Langhoff1988} The correct ordering of the triplet manifold was thus proposed to be a$^3\Pi$, b$^3\Sigma^+$, c$^3\Sigma^+$, d$^3\Pi$ and e$^3\Delta$, and this is the designation we will follow from now on. Of these, the lowest three states have been well characterized \cite{Truppe2019a,Walter2022,Doppelbauer2021,Walter2022b,Brown1978} but the d$^3\Pi$ state has never been reported upon and the rotational and fine-structure of the e$^3\Delta$ state, needed for its unambiguous characterization, has not been resolved. 

AlF has been of interest to spectroscopists all along: the electronic structure of AlF is similar to that of the intensively studied molecule CO, the benchmark diatomic molecule for studying perturbations.\cite{Hakalla} However, as AlF is heavier, its electronic transitions are in the more accessible region of the spectrum. When it became clear that the various properties of AlF make it an ideal candidate molecule for laser cooling and trapping experiments\cite{Truppe2019a,Hofsass2021,Wright2022}, e.g. because of its strong A$^1\Pi$ -- X$^1\Sigma^+$ transition with its highly diagonal Franck-Condon matrix\cite{DiRosa2004,Wells2011}, interest in the spectroscopic properties of AlF increased further. 

In the present study, we report on the missing d$^3\Pi$ state of AlF and experimentally characterize the e$^3\Delta$ state. In the early measurements of emission and absorption spectra of hot samples, the optical transitions connecting to these triplet states might have been obscured by other bands, although it is not a~priori clear why these states have not been observed earlier. We use multiple lasers to perform ionization spectroscopy in a jet-cooled molecular beam, and observe the $v=0-6$ levels of the d$^3\Pi$ state via excitation from laser-prepared, single rotational levels in the a$^3\Pi$ state and b$^3\Sigma^+$ state. Starting from the a$^3\Pi$ state, the $v=0-2$ levels of the e$^3\Delta$ state are characterized. In the 1974 overview paper\cite{Barrow1974a} the $v=0$ level of the next higher triplet state, named there the e$^3\Sigma^+$ state, has been reported upon as well. We characterize the $v=0-2$ levels of this same $^3\Sigma^+$ state, which we propose to refer to as the f$^3\Sigma^+$ state from now on. Apart from the energies of the lowest ro-vibrational levels of these three electronically excited triplet states, we determine their radiative lifetimes via time-delayed ionization. Ionization from the d$^3\Pi$ state allows to determine the absolute energies of the $v=0$ and $v=1$ levels in the X$^2\Sigma^+$ state of the AlF$^+$ cation, thereby significantly improving the accuracy of the ionization potential~(IP) relative to previous measurements.\cite{Dearden1991}

\section{Experiment} \label{sec:exp}

\begin{figure}[]
  \centering
   \includegraphics[width=240pt]{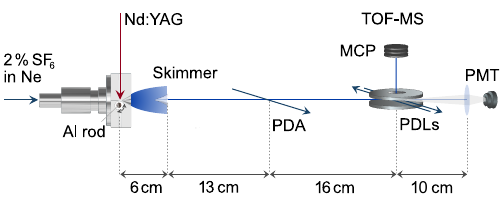}
   \caption{Scheme of the experimental setup. AlF molecules are produced by laser ablation of Al in a mixture of SF$_6$ seeded in neon. The molecules are cooled in the expansion and subsequently prepared in selected ro-vibrational levels of the a$^3\Pi$ state by a pulsed dye amplifier (PDA).
   Subsequently, the molecules are excited to higher lying triplet states and ionized by a pulsed dye laser (PDL). The AlF$^+$ ion signal is detected in a time-of-flight mass spectrometer (TOF-MS).}
    \label{fig:setup}
\end{figure}
The experimental setup is shown in Figure~\ref{fig:setup} and has been described in detail before\cite{Truppe2019a,Walter2022}. In short, a gas mixture of 2\,\% SF$_6$ in neon is released into vacuum through a pulsed solenoid valve (General Valve, Series 9) operating at 10\,Hz. Close to the valve opening, a rotating aluminum rod is laser-ablated by a focused Nd:YAG laser (Continuum Minilite I, \SI{1064}{\nano\meter}, \SI{16}{\milli\joule} pulse energy, \SI{5}{\nano\second} pulse duration, spot diameter \SI{0.5}{\milli\meter}). The ablated aluminum atoms react efficiently with SF$_6$ in a short reaction channel to form AlF molecules. These molecules are translationally and internally cooled through collisions with the carrier gas upon expanding from the channel into vacuum. Closely behind this laser ablation source, a pair of electrodes produces an electric field of 100\,V/cm, to deflect ions created in the laser ablation process. 

After collimation of the beam of neutral species through a conically shaped skimmer (\SI{4}{\milli\meter} diameter opening), the AlF molecules are excited to selected rotational levels in the metastable a$^3\Pi$ state, some \SI{19}{\centi\meter} downstream from the source. Excitation is performed on the $\Delta v=0$ bands of the a$^3\Pi, v$ $\leftarrow$ X$^1\Sigma^+, v$ transition around 367\,nm. For this, we use the frequency-doubled output of a pulsed dye amplifier (PDA), operated with Pyridine 2 dye, that is injection seeded by a narrow-band, cw  titanium sapphire laser (Sirah Matisse) and pumped by a frequency-doubled Nd:YAG laser (Innolas Spitlight, \SI{532}{\nano\meter}). The bandwidth of the frequency-doubled PDA radiation is about 220\,MHz. The PDA contains a stimulated Brillouin scattering (SBS) cell (to filter out amplified spontaneous emission) that induces a \SI{-1.98(5)}{\giga\hertz} shift to the fundamental frequency.\cite{Walter2022} The wavelength of the Ti:Sa seed-laser is measured with a calibrated wavemeter (HighFinesse WS8, \SI{10}{\mega\hertz} accuracy). At the experiment, the 367\,nm pulse energy is about 5\,--\,10\,mJ and the collimated PDA beam (5\,mm diameter) intersects the molecular beam perpendicularly. Because the lifetime of the metastable state is several milliseconds\cite{Walter2022}, the subsequent excitation and ionization steps can take place in a separate chamber, \SI{16}{\centi\meter} further downstream, where a a Wiley-McLaren type time-of-flight mass spectrometer is installed. In the field-free region between the repeller and extractor electrodes, the molecular beam is intersected with 1\,--\,\SI{300}{\micro\joule} of the frequency-doubled radiation of a pulsed dye laser (PDL, Sirah Cobra Stretch), tuned around \SI{280}{\nano\meter}. The spectral bandwidth of this laser is about 1.5\,GHz and its absolute frequency is recorded with a low-resolution wavelength meter (HighFinesse WS6-600, absolute accuracy \SI{600}{\mega\hertz}). The pulse energy is kept such as to avoid saturation and power broadening of the studied transitions. The UV excitation laser beam is spatially overlapped with \SI{7}{\milli\joule} of the counter-propagating beam from a second PDL (Radiant Dyes NarrowScan, operated on Rhodamine 101 dye). The wavelength of this laser is around \SI{620}{\nano\meter}, set to bring the AlF molecules above the ionization potential. The time delay between the excitation and ionization laser pulses is optimized for maximum signal. The voltages on the electrodes of the mass spectrometer are switched on some 100\,ns after the ionization laser pulse. The AlF$^+$ ions are then accelerated perpendicular to the molecular and laser beams, towards a microchannel plate detector. As the excitation and ionization of the AlF molecules takes place under field-free conditions, this detection scheme is parity selective. The ion signal is amplified and recorded on a computer using a fast digitizer card. The scans are controlled with home-built LabView data acquisition software.

We optimize the source conditions and the a$^3\Pi$ state preparation by detecting the a$^3\Pi$ $\rightarrow$ X$^1\Sigma^+$  phosphorescence signal by a photomultiplier tube (PMT; Hamamatsu R928). For this, a $f=50$\,mm fused silica imaging lens and a 367\,nm bandpass filter are installed at the end of the machine to collect radiation that is emitted along the molecular beam, i.e. in the forward direction.

\begin{figure}[]
        \centering
        \includegraphics[width=240pt]{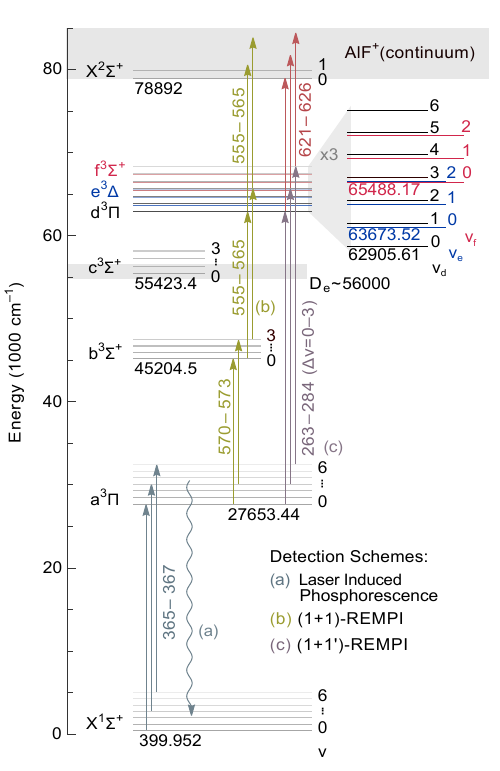}
        \caption{Energy level scheme of AlF, showing the ground state (X$^1\Sigma^+$), the lowest six triplet states (a$^3\Pi$\,--\,f$^3\Sigma^+$) and the ground state of the ion (X$^2\Sigma^+$).
        For each electronic state, the lowest vibrational levels are shown. The energies given in the plot are those of the $v=0$ level relative to the minimum of the X$^1\Sigma^+$ state potential.\cite{Barrow1974a} The vibrational levels shown on the expanded energy scale in the top right corner are those measured in the present study, which all lie above the dissociation energy ($D_e$), expected between 55000 and 56500\,cm$^{-1}$.
        The vertical arrows represent laser excitation, and are labeled by the relevant wavelength range in nm. Unless stated otherwise, diagonal bands ($\Delta v = 0$) are used for excitation. 
        }
        \label{fig:levelscheme}
        \end{figure}

Our first observation of the d$^3\Pi$ state was via excitation in the 555\,--\,565 \,nm region from selected ro-vibrational levels in the b$^3\Sigma^+$ state, followed by ionization from the upper state with the same laser. In this one-color resonance enhanced multiple photon ionization ((1+1)-REMPI) scheme, the pulse energy of the laser required for efficient ionization caused significant broadening of the resonant transitions from the b$^3\Sigma^+$ state. Nevertheless, the spectra were sufficiently resolved for the unambiguous assignment of the upper state as a normal $^3\Pi$ state with a value of the $A_v/B_v$ ratio of just below ten. The spectra also served for a first determination of the absolute energies of the lowest vibrational levels of this $^3\Pi$ state. The lowest ro-vibrational level of this $^3\Pi$ state is about 62507\,cm$^{-1}$ above the lowest ro-vibrational level in the X$^1\Sigma^+$ state of AlF. This is about 600\,cm$^{-1}$ higher than the predicted energy of the d$^3\Pi$ state.\cite{Langhoff1988} Given that the a$^3\Pi$ state, the b$^3\Sigma^+$ state and the c$^3\Sigma^+$ state have experimentally also been found some 740\,cm$^{-1}$, 570\,cm$^{-1}$ and 220\,cm$^{-1}$ higher in energy, respectively, than in these theoretical calculations\cite{Langhoff1988}, we can safely conclude that we deal with the sought-after d$^3\Pi$ state. 

For a more detailed characterization of the d$^3\Pi$ state we used two-color (1+1')-REMPI, starting from selected laser-prepared ro-vibrational levels in the a$^3\Pi$ state. This two-color scheme enables the independent optimization of the resonant excitation and ionization steps. By using the a$^3\Pi$ state as the intermediate state, both the e$^3\Delta$ and the f$^3\Sigma^+$ states that are expected in close proximity to the d$^3\Pi$ state can be reached and investigated as well. By scanning the time-delay between the excitation and the ionization laser the radiative lifetime of the levels in either one of these triplet states can be measured. Moreover, by scanning the wavelength of the ionization laser, the ionization potential (IP) can be accurately determined.

In these experiments, the AlF molecules are excited on the R$_2(N)$ ($N=0-4$) lines of the $\Delta v=0$ bands of the a$^3\Pi_1, v$\,--\,X$^1\Sigma^+, v$ transition. This way, we selectively populate the $J=N+1$ rotational level with $(-1)^J$ parity in the a$^3\Pi_1, v$ state. We perform these experiments starting from $v=0-6$ in the X$^1\Sigma^+$ state; in the molecular beam, the population in the $v=6$ level is about two orders of magnitude less than in the $v=0$ level, and beyond that it is challenging to obtain an adequate signal-to-noise ratio. We then scan the 280\,nm UV excitation laser over a range of 20\,--\,30\,cm$^{-1}$ to map out the ro-vibrational levels in the d$^3\Pi$, e$^3\Delta$ and f$^3\Sigma^+$ states that can be reached from this laser-prepared level in the a$^3\Pi$ state. 
In total, we record in total the absolute energies of 183 ro-vibrational levels, that are all listed in the Supporting Information. 

\section{Theory} \label{sec:theory}

The general Hamiltonian to describe the energy level structure in the electronic states of AlF can be expressed as
\begin{equation}
    H=H_{\text{ev}}+H_{\text{rot}}+H_{\text{fs}}+H_{\text{hfs}}
\end{equation}
but we can neglect the last term, due to the hyperfine structure, as this has not been experimentally resolved in the present study.
The term $H_{\text{ev}}$ describes the electronic and vibrational energy $E_{v}$ that is commonly expressed as
\begin{eqnarray}
    E_{v}(v)=&&T_e+\omega_e\left(v+\frac{1}{2}\right)-\omega_e x_e\left(v+\frac{1}{2}\right)^2
\end{eqnarray}
We follow the convention of Barrow {\it et~al.}\cite{Barrow1974a}, and fix $T_e=0$ for the minimum of the X$^1\Sigma^+$ ground state potential. This puts the lowest ro-vibrational level in the X$^1\Sigma^+$ state at 399.952\,cm$^{-1}$, as indicated in Figure~\ref{fig:levelscheme}.
The rotational part of the Hamiltonian, $H_{\text{rot}}$, and the fine-structure part, $H_{\text{fs}}$, is best expressed in terms of the Hund's case~(a) model for the a$^3\Pi$ state. We therefore describe both the d$^3\Pi$ state and the e$^3\Delta$ state in terms of Hund's case~(a) as well, even though, as we will see, the ratio $A_v/B_v$ of the spin-orbit coupling constant ($A$) to the rotational constant ($B$) is much smaller in these states than in the a$^3\Pi$ state. The expression for $H_{\text{rot}}$ + $H_{\text{fs}}$ that we use is given by
\begin{equation}
H_{\text{rot}} + H_{\text{fs}} =A_v \, L_z \,S_z + \frac{2}{3}\lambda_v (3S_z^2 - \mathbf{S}^2)+B_v(\mathbf{J}-\mathbf{L}-\mathbf{S})^2
\end{equation}
where $\mathbf{L}$ and $\mathbf{S}$ are the total electron angular momentum and total electron spin, respectively, and where $\mathbf{J}$ is the total angular momentum, including the end-over-end rotation $\mathbf{N}$. The projections of $\mathbf{L}$ and $\mathbf{S}$ on the internuclear axis are $L_z$ and $S_z$, respectively, and $\lambda$ is the spin-spin interaction parameter. The sub-indices refer to the specific vibrational level $v$ that these parameters belong to, and for the vibrational dependence of $A_v$ and $B_v$ we use the expressions
\begin{align}
A_v&= A_e - \zeta_e \left(v+\frac{1}{2}\right) \\
B_v&=B_e-\alpha_e\left(v+\frac{1}{2}\right)
\end{align}
The f$^3\Sigma^+$ state is best described as Hund's case~(b) and the energy of the low rotational levels that we measured is given by $E_{\text{rot}} = B_v N(N+1)$. Some broadening due to spin-rotation, spin-spin or hyperfine interactions is discernible in the spectral lines to this state, but the spectral resolution is insufficient to analyse this further.

\section{Results} 
\subsection{The d$^3\Pi$ state of AlF} \label{sec:res-d}

\begin{figure}[]
        \centering
        \includegraphics[width = 240pt]{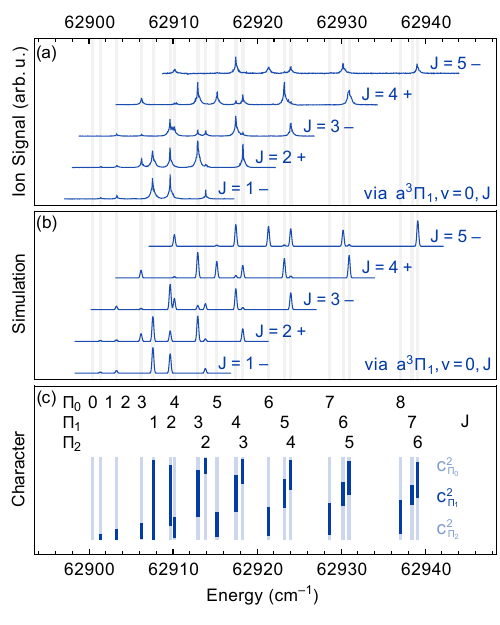}
        \caption{(a) Rotationally resolved spectra to the d$^3\Pi$, $v=0$ state recorded via different rotational levels $J$ in the a$^3\Pi_1,v=0$ state. The individual traces are normalized and vertically offset for clarity. The energy scale is relative to the minimum of the X$^1\Sigma^+$ state potential. (b) Simulated spectra, see Section~\ref{sec:res-d} for details. (c) Vertical bars, indicating the fraction of $\Omega=0, 1$ (in bold) and 2 character of the $J$-levels in the d$^3\Pi, v=0$ state.}
        \label{fig:v0_full}
\end{figure}

To investigate the $v=0-6$ levels in the d$^3\Pi$ state, it is experimentally convenient to use excitation on the $\Delta v=0$ bands of the d$^3\Pi, v$ -- a$^3\Pi, v$ transition as these are spectrally rather close together. Figure~\ref{fig:v0_full}a shows the experimental spectra to the $v=0$ level in the d$^3\Pi$ state, obtained via the $v=0$, $J=1-5$ levels in the a$^3\Pi_1$ state. The spectra are plotted on an absolute energy scale, using the known energies of the ro-vibrational levels in the X$^1\Sigma^+$ state\cite{Hedderich1992} and the measured energies of the two excitation lasers. The energy levels in the d$^3\Pi, v=0$ state can be readily assigned a $J$ quantum number, as in Figure~\ref{fig:v0_full}c. The vertical bars in Figure~\ref{fig:v0_full}c indicate the fraction of $\Omega=0, 1$ and 2 character of each $J$-level; the fraction of $\Omega=1$ character is indicated in bold, and the fraction of $\Omega=0$ and $\Omega=2$ character is in lighter color above and below that, respectively. In the spectrum shown in Figure~\ref{fig:v0_full}a, recorded via the $J=1$ level in the a$^3\Pi_1$ state, the observed five peaks can be assigned, from low to high energy, to $\Omega=0,J=1,2$; $\Omega=1,J=1,2$; $\Omega=2,J=2$, where the $\Omega$ labeling is only as good as given by the bars in Figure~\ref{fig:v0_full}c. The transition to $\Omega=0,J=0$ is very weak and only visible under excitation conditions at which the other spectral lines are strongly saturated. The same $J$-levels in the d$^3\Pi$ state are reached via excitation from rotational levels with either plus or minus parity in the a$^3\Pi_1$ state. As no shift between transitions from plus and minus parity levels is observed, we conclude that the $\Lambda$-doublet splitting in the d$^3\Pi$ state is considerably smaller than the width of the lines in the spectra.

\begin{table} \centering
  \caption{Energies $E_{v}$, rotational constants $B_v$, spin-orbit and spin-spin coupling constants $A_v$ and $\lambda_v$ of the experimentally observed vibrational levels $v$ in the d$^3\Pi$ state. The parameter $\sigma$ is the standard deviation of the fit.
  All values are given in cm$^{-1}$.}
  \label{table:dstaterotconst}
  \begin{tabular}{ llllll }
   \hline
 $v$ & $E_{v}$ & $B_v$ & $A_v$ & $\lambda_v$ & $\sigma$\\
 \hline
 0 &62905.61(5)~ & 0.591(2) & 5.74(3) & $-$0.02(4) & 0.03\\
 1 &63839.47(6)~ & 0.585(2) & 5.75(4) & 0.02(4) & 0.03\\
 2 &64763.15(6)~ & 0.580(2) & 5.80(3) & 0.03(4) & 0.03\\
 3 &65676.63(6)~ & 0.576(2) & 5.81(3) & 0.06(4) & 0.03\\
 4 &66579.45(3)~ & 0.571(1) & 5.84(2) & 0.05(2) & 0.02\\
 5 &67471.92(9)~ & 0.566(3) & 5.84(5) & $-$0.02(5) & 0.04\\
 6 &68353.67(17) & 0.560(6) & 5.89(9) &  0.03(10) & 0.09\\
  \hline
\end{tabular}
\end{table}

From the set of measurements on the $v=0-6$ levels of the d$^3\Pi$ state, the rotational constants, the spin-orbit coupling constants and the spin-spin coupling constants have been determined and these are given in Table~\ref{table:dstaterotconst}. As stated earlier, the absolute energies are given relative to the minimum of the potential of the X$^1\Sigma^+$ state.\cite{Barrow1974a} This places the $J=1$ level of the $\Omega=1$ manifold of the d$^3\Pi, v=0$ state 62507.69 cm$^{-1}$ above the energy of the $N=0$, $v=0$ level in the X$^1\Sigma^+$ ground state.  Figure~\ref{fig:v0_full}b shows the simulated spectra that correspond to the experimental spectra in Figure~\ref{fig:v0_full}a. 
The line intensities are calculated by 
\begin{equation}
   I \propto  \left(   c_{\Pi_1}\left(\Omega_{\text{d}},J_{\text{d}}\right) 
 \, \mathcal{S}_{\text{ad}} (v_{\text{a}},v_{\text{d}})\right)^2 \mathcal{L}_{^3\Pi\shortleftarrow^3\Pi},
   \label{eq:Ipi-pi}
\end{equation}
where $|c_{\Pi_1}|^2$ is the amount of $\Pi_1$ character of the reached level in the d$^3\Pi$ state, derived from the fitted spectroscopic parameters as given in Table~\ref{table:dstaterotconst} and listed in the Supporting Information.
$\mathcal{S}_{\text{ad}}(v_{\text{a}},v_{\text{d}})$ is the overlap integral of the vibrational wave functions, and $\mathcal{S}^2_{\text{ad}}(v_{\text{a}},v_{\text{d}})$ the according Franck-Condon factors, as given in Table~\ref{tab:fc}.
The H\"onl-London are those of a $^3\Pi$\,$\leftarrow$\,$^3\Pi$ transition, where the lower state is considered a pure Hund's case~(a), and are given by\cite{Budo1937}: 
\begin{align}
\mathcal{L}_{^3\Pi\shortleftarrow^3\Pi}^{ \text{P}_2(J'')} & = (J''-1)(J''+1)/J''\\
\mathcal{L}_{^3\Pi\shortleftarrow^3\Pi}^{ \text{Q}_2(J'')}  & =  (2J''+1)/ \left( J''(J''+1) \right) \\
\mathcal{L}_{^3\Pi\shortleftarrow^3\Pi}^{ \text{R}_2(J'')} &  = J''(J''+2)/ \left( J''+1 \right) 
\end{align}
while all others are zero. 

 \begin{figure}[]
        \centering
        \includegraphics[width = 240pt]{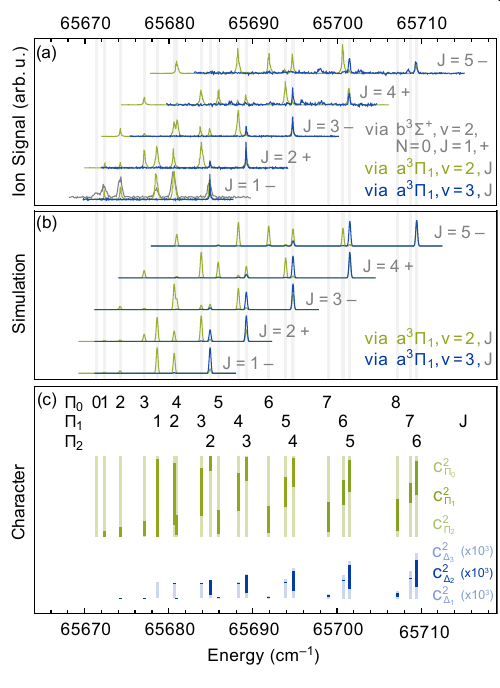}
      \caption{(a) Rotationally resolved spectra to the d$^3\Pi$, $v=3$ state recorded via different rotational levels $J$ in the a$^3\Pi_1,v=2$ state (green traces) and in the a$^3\Pi_1,v=3$ state (blue traces). The gray trace is recorded using one-color (1+1)-REMPI from the b$^3\Sigma^+, v=3, N=0$, $J=1$ level with positive parity. The individual traces are normalized and vertically offset for clarity.
      (b) Simulated spectra, see Section~\ref{sec:res-pert} for details. (c) The upper row of vertical bars, in green, indicate the fraction of $\Omega=0, 1$ (in bold) and 2 character of the $J$-levels in the d$^3\Pi, v=3$ state. The lower row of vertical bars, in blue, indicate the fraction of $\Delta_1$, $\Delta_2$ (in bold) and $\Delta_3$ character of the $J$-levels in the d$^3\Pi, v=3$ state due to mixing with the e$^3\Delta, v=2$ state.
      }
        \label{fig:v3_full}
\end{figure}

The spectra recorded for the diagonal bands of the d$^3\Pi, v$ -- a$^3\Pi, v$ transition all appear similar, apart from those belonging to the d$^3\Pi, v=3$ -- a$^3\Pi, v=3$ band. The latter spectra are rather weak and unexpectedly show a drastically different rotational intensity pattern, with mainly the lines to the rotational levels in the $\Omega=2$ manifold being detectable, as shown by the blue traces in the Figure~\ref{fig:v3_full}a. The set of spectra to the $v=3$ level in the d$^3\Pi$ state recorded via the same set of rotational levels in the a$^3\Pi_1, v=2$ state, i.e. on an off-diagonal band, are shown as the green traces in the Figure~\ref{fig:v3_full}a. Although these spectra are slightly saturated, they show the normal, expected line positions and intensity pattern and are very similar to the spectra shown in Figure~\ref{fig:v0_full}. From these observations it is concluded that the transition dipole moment of the d$^3\Pi, v=3$ -- a$^3\Pi, v=3$ band is near zero, and that the intensity observed in this band results from the intensity borrowing from a nearby transition. This intensity borrowing can occur when the d$^3\Pi, v=3$ state is perturbed by another state, to which a strong transition from the a$^3\Pi, v=3$ state exists.\cite{James1971} It will be shown in Section~\ref{sec:res-pert}, that this intensity borrowing results from the weak perturbation of the d$^3\Pi$ state with the e$^3\Delta$ state. At first, this perturbation does not appear to result in a detectable shift or distortion of the rotational energy levels in $v=3$, nor in any of the other vibrational levels of the d$^3\Pi$ state. 

We measured the lifetime of several low-$J$ levels in the d$^3\Pi, v=0, 1$ and 2 states, and always find values of $(40 \pm 4)$~ns. This agrees well with the calculated lifetime of the d$^3\Pi$ state of 48.2~ns.\cite{Langhoff1988} 

\subsection{The e$^3\Delta$ state of AlF} \label{sec:res-e}

\begin{figure}[]
        \centering
        \includegraphics[width = 240pt]{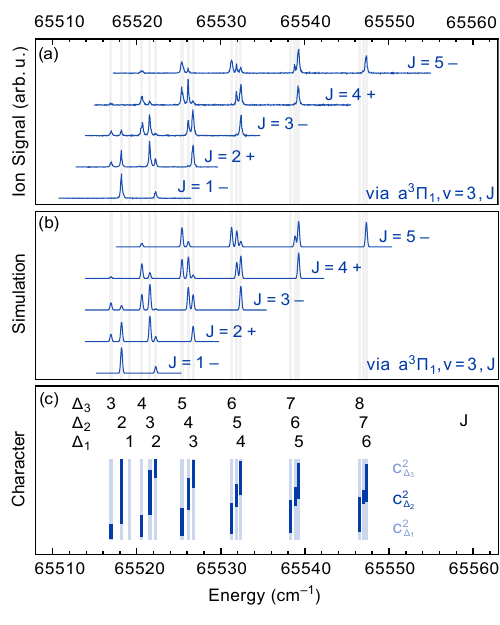}
        \caption{(a) Rotationally resolved spectra to the e$^3\Delta$, $v=2$ state recorded via different rotational levels $J$ in the a$^3\Pi_1,v=3$ state. The individual traces are normalized and vertically offset for clarity. (b) Simulated spectra, see Section~\ref{sec:res-e} for details.  (c) Vertical bars, indicating the fraction of $\Omega=1, 2$ (in bold) and 3 character of the $J$-levels in the e$^3\Delta, v=2$ state. Note that this state is inverted.}
        \label{fig:ev2_full}
        \end{figure}

As mentioned in the Introduction, the lowest two vibrational levels of a state tentatively assigned as a $^3\Delta$ state\cite{So1974} have been observed earlier, about 36018\,cm$^{-1}$ and 36948\,cm$^{-1}$, respectively, above the $v=0$ level of the a$^3\Pi$ state.\cite{Dodsworth1955} This places these levels about 160\,cm$^{-1}$ below the $v=1$ and $v=2$ levels, respectively, of the d$^3\Pi$ state. In order to characterize the lowest vibrational levels of the e$^3\Delta$ state, we scanned the spectral region down to 200\,cm$^{-1}$ to the red from the diagonal d$^3\Pi, v$ -- a$^3\Pi, v$ bands. The spectra of the e$^3\Delta, v=2$ -- a$^3\Pi, v=3$ band, plotted on an absolute energy scale, are shown in Figure~\ref{fig:ev2_full}a. The spectra recorded via the low-$J$ levels in the a$^3\Pi$ state enable an unambiguous assignment of the upper state as a $^3\Delta$ state with a small, negative $A_v/B_v$ ratio, that is, this state is close to Hund's case~(b) and has an inverted spin structure.

The spectra to the $v=0$ and $v=1$ levels of the e$^3\Delta$ state appear similar to those shown in Figure~\ref{fig:ev2_full}a, and the rotational constants and spin-orbit coupling constants deduced from these spectra are given in Table~\ref{table:estaterotconst}. Using these values, the $J=2$ level of the $\Omega=2$ manifold of the e$^3\Delta, v=0$ state is found 63274.98 cm$^{-1}$ above the energy of the $N=0$, $v=0$ level in the X$^1\Sigma^+$ ground state.   

It is understandable from the appearance of the energy level structure shown in Figure~\ref{fig:ev2_full}c, that this upper state has been mistaken for a $^3\Sigma$ state in the early days. Only the observation of the lines to the lowest rotational levels makes the labeling with the $\Omega$ and $J$ quantum numbers as indicated in Figure~\ref{fig:ev2_full}c possible. The vertical bars in this panel indicate the fraction of $\Omega=1, 2$ and 3 character of each $J$-level; the fraction of $\Omega=2$ character is indicated in bold, and the fraction of $\Omega=1$ and $\Omega=3$ character is in lighter color below and above that, respectively.

Figure~\ref{fig:ev2_full}b shows the simulated spectra that correspond to the experimental spectra shown in the panel above. 
The line intensities are calculated by 
\begin{equation}
   I \propto  \left(   c_{\Delta_2}\left(\Omega_{\text{e}},J_{\text{e}}\right) 
 \, \mathcal{S}_{\text{ae}} (v_{\text{a}},v_{\text{e}})\right)^2 \mathcal{L}_{^3\Delta\shortleftarrow^3\Pi},
   \label{eq:Idelta-pi}
\end{equation}
where $|c_{\Delta_2}|^2$ is the amount of $\Delta_2$ character of the reached level in the e$^3\Delta$ state, derived from the fitted spectroscopic parameters as given in Table~\ref{table:estaterotconst} and listed in the Supporting Information.
$\mathcal{S}_{\text{ae}} (v_{\text{a}},v_{\text{e}})$ is the overlap integral of the vibrational wave functions, and $\mathcal{S}_{\text{ae}}^2 (v_{\text{a}},v_{\text{e}})$ the according Franck-Condon factors, as given in Table~\ref{tab:fc}.
The H\"onl-London are those of a $^3\Delta$\,$\leftarrow$\,$^3\Pi$ transition, where the lower state is considered a pure Hund's case~(a), and read as\cite{Kovacs1963}: 
\begin{align}
\mathcal{L}_{^3\Delta\shortleftarrow^3\Pi}^{ \text{P}_2(J'')}  &= (J''-2)(J''-1)/J''\\
\mathcal{L}_{^3\Delta\shortleftarrow^3\Pi}^{ \text{Q}_2(J'')}  &=  (J''-1)(J''+2)(2J''+1)/\left(J''(J''+1) \right) \\
\mathcal{L}_{^3\Delta\shortleftarrow^3\Pi}^{ \text{R}_2(J'')}  &=  (J''+2)(J''+3)/\left(J''+1 \right)
\end{align}
while all others are zero. 
The experimental spectra are seen to be well reproduced by the simulations. The e$^3\Delta$$_1, J=1$ level cannot be reached from the a$^3\Pi_1,J=1$ level as the H\"onl-London factors for $^3\Delta_{\Omega}$ -- $^3\Pi_1$ transitions are only non-zero for $\Omega=2$. We can detect this level by exciting the ground state molecules to the a$^3\Pi_0,J=0$ level on the P$_1$(1) transition before driving the e$^3\Delta$ $\leftarrow$ a$^3\Pi_0$ transition.

\begin{table} \centering
  \caption{Energies $E_{v}$, rotational constants $B_v$ and  spin-orbit  coupling constants $A_v$ of the experimentally observed vibrational levels $v$ in the e$^3\Delta$ state. The parameter $\sigma$ is the standard deviation of the fit.
  All values are given in cm$^{-1}$.}
  \label{table:estaterotconst}
  \begin{tabular}{ rllll }
  \hline
 $v$ & $E_v$ & $B_v$ & $A_v$  & $\sigma$\\
 \hline
  0 & 63673.52(6) & 0.589(2) & $-$0.41(1) &  0.02\\
 1 & 64601.70(6) & 0.584(2) & $-$0.46(1) &  0.02\\
 2 & 65516.7(1) & 0.580(3) & $-$0.62(2) &  0.04\\
 \hline
\end{tabular}
\end{table}

The intensities of the strong lines in the spectra of the e$^3\Delta, v=2$ -- a$^3\Pi, v=3$ band are orders of magnitude larger than those of the nearby d$^3\Pi, v=3$ -- a$^3\Pi, v=3$ band. If we assume that the ionization efficiencies from the e$^3\Delta, v=2$ and d$^3\Pi, v=3$ states are the same, we can estimate the intensity ratio of the two bands by comparing the excitation pulse energies required for the same ion signal. This puts the ratio of the band intensities at $(2.5 \pm 0.5) \times 10^{-4}$. 

The calculated lifetime of the e$^3\Delta$ state is 6.0\,ns.\cite{Langhoff1988} Such a lifetime is too short to be accurately determined via time-delayed ionization with laser pulses of 5\,ns duration. Our time-delayed ionization measurements confirm that the lifetime of several low-$J$ levels in the e$^3\Delta, v=2$ state is very short, certainly below 10~ns.

\subsection{The $\text{f}^3\Sigma^+$ state of AlF} \label{sec:res-f}

\begin{figure}[]
        \centering
        \includegraphics[width = 240pt]{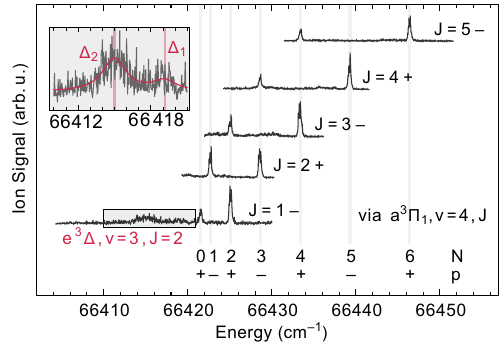}
        \caption{Rotationally resolved spectra to the f$^3\Sigma^+$$, v=1$ state recorded via different rotational levels $J$ in the a$^3\Pi_1,v=4$ state. The individual traces are normalized and vertically offset for clarity. The broad structure at around 66415\,cm$^{-1}$ in the lowest trace (shown in the inset) is due to transitions to the two $J=2$ levels of the predissociated e$^3\Delta,v=3$ state. 
        } 
        \label{fig:fv1_full}
\end{figure}

Some 30\,cm$^{-1}$ below the e$^3\Delta, v=2$ -- a$^3\Pi, v=3$ band, we observe transitions from the a$^3\Pi, v=3$ state to yet another electronically excited state. This state is readily identified as a $^3\Sigma^+$ state, as only levels with $N=J \pm 1$ can be reached when exciting from $J$ levels in the a$^3\Pi_1$ state with parity $(-1)^J$. We refer to this state as the f$^3\Sigma^+, v=0$ state, even though this state has previously been documented as the e$^3\Sigma^+, v=0$ state.\cite{Barrow1974a} Our measurements reproduce the earlier reported term values, and in addition we observe the $N=0$ level. We explicitly verify that there is no lower vibrational level belonging to this f$^3\Sigma^+$ state, and we do identify its $v=1$ and $v=2$ levels. The spectra of the f$^3\Sigma^+, v=1$ -- a$^3\Pi_1, v=4$ band are shown in Figure~\ref{fig:fv1_full}. It is seen that the $R(J)$ lines are always stronger than the $P(J)$ lines. A triplet structure is observed on the rotational lines of the f$^3\Sigma^+$ $\leftarrow$a$^3\Pi_1$ transition, that appears similar to that observed on the b$^3\Sigma^+$ $\leftarrow$ a$^3\Pi$ transition, where the hyperfine structure has been fully resolved.\cite{Doppelbauer2021}

The energies and rotational constants that we extract from these spectra are given in Table~\ref{table:fstaterotconst}. In this case, $E_{v}$ is the energy of the $N=0$ level in the f$^3\Sigma^+, v$ state. It should be noted that the lines to the f$^3\Sigma^+, v=2$ state appear around twice as broad than those to the lower vibrational states, which might be indicative of predissociation.

The broad structure on the baseline that can be observed in several traces shown in Figure~\ref{fig:fv1_full} is attributed to residual rotational structure of the e$^3\Delta, v=3$ -- a$^3\Pi_1, v=4$ band. When reducing the intensity of the excitation laser, this broad structure becomes relatively more pronounced indicating that it originates from a band that is stronger than the f$^3\Sigma^+, v=1$ -- a$^3\Pi, v=4$ band. Based on the extrapolation from the $v=0-2$ data, the two $J=2$ levels of the e$^3\Delta, v=3$ state are expected just above 66420\,cm$^{-1}$. Two broadened lines, with the expected spacing of about 4.1\,cm$^{-1}$ and with the expected intensity ratio can be discerned when exciting from the $J=1$, negative parity level in the a$^3\Pi_1, v=4$ state, as shown in the inset to Figure~\ref{fig:fv1_full}. It appears that the rotational levels in the e$^3\Delta, v=3$ state are down-shifted by about 5\,cm$^{-1}$ and that the rotational lines exciting these levels are lifetime-broadened to about 1.3\,cm$^{-1}$, indicating predissociation on a time-scale of about 4\,ps.

\begin{table} \centering
  \caption{Energies $E_{v}$ and rotational constants $B_v$ of the experimentally observed vibrational levels $v$ in the f$^3\Sigma^+$ state. The parameter $\sigma$ is the standard deviation of the fit.
  All values are given in cm$^{-1}$.}
  \label{table:fstaterotconst}
 \begin{tabular}{ rlll }
 \hline
$v$ & $E_{v}$ & $B_v$   & $\sigma$\\
 \hline
 0 & 65488.17(4) & 0.594(1) &   0.02 \\
 1 & 66421.50(6) & 0.593(1) &  0.03  \\
 2 & 67343.16(8) & 0.582(2) &  0.07\\
 \hline
\end{tabular}
\end{table}

We measured the lifetime of several low-$J$ levels in the f$^3\Sigma^+, v=0$ state, and find values of $(30 \pm 6)$~ns. This is comparable to the lifetime of the d$^3\Pi$ state. Both the f$^3\Sigma^+$ and the d$^3\Pi$ state can decay with a significant rate to the c$^3\Sigma^+$, the b$^3\Sigma^+$ and the a$^3\Pi$ state. 

In Table~\ref{table:vibconsts}, the equilibrium parameters for the electronic potentials of the d$^3\Pi$, e$^3\Delta$ and f$^3\Sigma^+$ states of AlF are given,
as well as the radiative lifetime $\tau$. The equilibrium internuclear distances $r_\text{{eq}}$ are the values determined from the equilibrium rotational constants given above, reduced by 0.0027\,\AA~for the d$^3\Pi$ state and increased by 0.0027\,\AA~for the e$^3\Delta$ state (see Section~\ref{sec:res-pert}).

\begin{table} 
\centering
  \caption{Equilibrium parameters for the electronic potentials of the d$^3\Pi$, e$^3\Delta$ and f$^3\Sigma^+$ states of AlF.
    If not stated otherwise, all values are given in cm$^{-1}$.}
  \label{table:vibconsts}
\begin{tabular}{ l l l l }
 \hline 
  & d$^3\Pi$ & e$^3\Delta$ & f$^3\Sigma^+$ \\
 \hline
$T_e$ &62434.6(3) & 63204.5(3)& 65017.1(3)\\
$\omega_e$ & 944.5(2) & 941.3(1) & 945.0(1) \\
$\omega_ex_e$ & 5.21(4) & 6.57(10) & 5.84(10) \\
$B_{e,\text{eff}} $ & 0.5931(4)  & 0.5911(5) & 0.599(5) \\
$\alpha_e$ & 0.0050(1) & 0.0045(3) & 0.006(3) \\
$A_e$ & 5.73(1) & $-$0.34(5) & $-$ \\
$\zeta_e$ & $-$0.024(3) & 0.11(3) & $-$ \\
\hline
$\tau$ (ns)& 40(4) & $<$ 10 & 30(6) \\  
\hline
$r_\text{{eq}}$ (\AA) &
1.5940$^{a}$ & 1.6021$^{a}$ & 1.5888 \\
$B_{e,\text{depert}} $ & 0.5951$^{b}$  & 0.5891$^{b}$ & $-$ \\
 \hline
\end{tabular}\\
\textsuperscript{\emph{a}} Derived from $B_{e,\text{eff}}$ after deperturbation.\\
\textsuperscript{\emph{b}} Calculated from $r_\text{{eq}}$.
\end{table}

\begin{table} \centering
  \caption{Franck-Condon factors $\mathcal{S}_{\text{ad}}^2$, $\mathcal{S}_{\text{ae}}^2$ and $\mathcal{S}_{\text{af}}^2$ of the
   a$^3\Pi$--d$^3\Pi$, 
 a$^3\Pi$--e$^3\Delta$ and 
   a$^3\Pi$--f$^3\Sigma^+$ bands, calculated using a Morse potentials derived from the parameters $\omega_e$, $\omega_ex_e$ and $r_{\text{eq}}$ as given in Table~\ref{table:vibconsts}.
For the corresponding values of the  a$^3\Pi$ state, see Ref. \cite{Walter2022}.
  }
  \label{tab:fc}
  \begin{tabular}{lllllll}
    \hline
 &   $v_\text{a}=0$ &     $v_\text{a}=1$ &   $v_\text{a}=2$ &   $v_\text{a}=3$ &   $v_\text{a}=4$ & $v_\text{a}=5$  \\
    \hline 
$v_\text{d}=0$ & 0.661 & 0.264 & 0.062 & 0.011 & 0.002 & 0.000 \\ 
$v_\text{d}=1$ & 0.282 & 0.227 & 0.318 & 0.132 & 0.034 & 0.007 \\ 
$v_\text{d}=2$ & 0.052 & 0.369 & 0.043 & 0.269 & 0.182 & 0.065 \\ 
$v_\text{d}=3$ & 0.005 & 0.122 & 0.347 & 0.000 & 0.185 & 0.204 \\ 
$v_\text{d}=4$ & 0.000 & 0.018 & 0.189 & 0.275 & 0.024 & 0.105 \\ 
$v_\text{d}=5$ & 0.000 & 0.001 & 0.038 & 0.242 & 0.190 & 0.070 \\ 
\hline
$v_\text{e}=0$ & 0.749 & 0.212 & 0.035 & 0.004 & 0.000 & 0.000 \\ 
$v_\text{e}=1$ & 0.222 & 0.384 & 0.297 & 0.081 & 0.014 & 0.002 \\ 
$v_\text{e}=2$ & 0.028 & 0.331 & 0.175 & 0.309 & 0.125 & 0.028 \\ 
$v_\text{e}=3^\text{a}$ & 0.002 & 0.068 & 0.370 & 0.064 & 0.282 & 0.159 \\ 

\hline
$v_\text{f}=0$ & 0.611 & 0.288 & 0.081 & 0.017 & 0.003 & 0.001 \\ 
$v_\text{f}=1$ & 0.315 & 0.159 & 0.307 & 0.156 & 0.049 & 0.012 \\ 
$v_\text{f}=2^\text{a}$ & 0.067 & 0.378 & 0.013 & 0.226 & 0.196 & 0.086 \\ 
 
\hline
  \end{tabular}\\
  \textsuperscript{\emph{a}} Predissociated
\end{table}

\subsection{The $\text{X}^2\Sigma^+$ state of AlF$^+$} \label{sec:res-Ion}

The ro-vibrational levels of the d$^3\Pi$ state are ideally suited as intermediate levels for an accurate measurement of the ionization potential (IP) of AlF. These levels are only one visible photon away from the IP, and have a sufficiently long lifetime to enable a well-defined, single-photon ionization experiment. For this, the AlF molecules are prepared in a single ro-vibrational level of the d$^3\Pi, v=0$ state. After a time-delay of about 20~ns, the ionization laser is fired to excite the molecules further. The ionization laser is scanned over the region where the ionization onset is expected. Laser preparation and ionization take place in zero electric field, but to detect the ions the extraction fields of the mass spectrometer need to be switched on at some point. These extraction fields cause field-ionization of laser-prepared, long-lived Rydberg states. Instead of seeing a sharp ionization onset, the onset of ionization will occur gradually with increasing photon energy, and begins at energies below what is required to reach the lowest rotational levels in the cation. It is therefore important to wait as long as possible with switching on the extraction fields such that the Rydberg states have decayed. The geometry of our ionization detection region and the beam velocity of about 800\,m/s set an upper limit to this wait-time of about 12\,$\mu$s.

\begin{figure}[]
        \centering
        \includegraphics[width = 240pt]{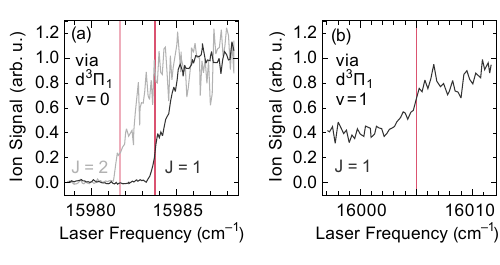}
        \caption{
         Measurements of the  AlF$^+$ ion signal versus ionization laser energy.
         (a) The onset of ionization from the $J=1$ (black) and $J=2$ (gray) level of the d$^3\Pi_1, v=0$ state is used to determine the ionization potential of AlF.
         (b) The increase in ionization signal starting from the d$^3\Pi_1, v=1,J=1$ level of AlF occurs as ionization to the first vibrationally excited state in the ion becomes possible.
        }
        \label{fig:IP-onset}
\end{figure}

Measurements of the ion signal as a function of the photon energy of the ionization laser, with a pulsed-field extraction delay of 12\,$\mu$s, are shown in Figure~\ref{fig:IP-onset}. In Figure~\ref{fig:IP-onset}a, two ionization onset curves, starting from the $J=1$ level (higher onset) and from the $J=2$ level (lower onset) of the d$^3\Pi_1, v=0$ state are shown. The onsets are seen to occur at frequencies that differ by the energy separation of the starting levels in the d$^3\Pi$ state, as expected. Moreover, in both curves a clear step can be recognized, that can be attributed to first reaching only the $N=0$ level and then, at about 1.2\,cm$^{-1}$ higher energy, also the $N=1$ level of the X$^2\Sigma^+$ state of the AlF$^+$ cation. From these measurements, the energy of the $N=0$ level of the X$^2\Sigma^+, v=0$ state of the AlF$^+$ cation is concluded to be at least 78491\,cm$^{-1}$ but not more than 78493\,cm$^{-1}$ above the $N=0$ level of the X$^1\Sigma^+, v=0$ state of AlF. It is difficult to define the field-free IP more precisely than this, as stray electric fields of 10\,mV/cm can already lower the IP by 0.5\,cm$^{-1}$. This value of $(78492 \pm 1$)\,cm$^{-1}$ for the IP is about 20\,cm$^{-1}$ larger, and over one order of magnitude more accurate, than the best value reported for this to date.\cite{Dearden1991}

In Figure~\ref{fig:IP-onset}b a step in the ionization signal, increasing the ion signal by about a factor two, is seen in the same spectral region when starting from the $J=1$ level of the d$^3\Pi_1, v=1$ state. This step results from reaching the X$^2\Sigma^+, v=1$ state of the cation. Relative to the onset starting from the $J=1$ level of the d$^3\Pi_1, v=0$ state shown in Figure~\ref{fig:IP-onset}a, this step is about $(21 \pm 2)$\,cm$^{-1}$ higher in energy. This means that $\omega_e-2\omega_e x_e$ in the X$^2\Sigma^+$ state of the ion is $(955 \pm 2)$\,cm$^{-1}$, which compares well to the calculated value of 958\,cm$^{-1}$.\cite{Kang_2017}

\subsection{Perturbation of the $\text{d}^3\Pi$ and $\text{e}^3\Delta$ states} \label{sec:res-pert}

\begin{figure}[]
        \centering
        \includegraphics[width=450pt]{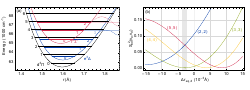}
        \caption{ (a) Expanded view of the Morse potentials of the d$^3\Pi$, e$^3\Delta$ and f$^3\Sigma^+$ states around their minima. The  vibrational levels that have been observed for each of these states are indicated.
        Dashed lines schematically illustrate potential barriers for the states where predissociation is observed.
    (b) Calculated Franck-Condon factors $\mathcal{S}_{\text{ad}}^2(v_{\text{a}},v_{\text{d}})$  for the diagonal bands of the d$^3\Pi$ -- a$^3\Pi$ transition as a function of $\Delta r_{\text{eq,d}}$, which is the difference of the equilibrium internuclear distance of the d$^3\Pi$ state from 1.5967\,\AA, the value obtained from the effective (perturbed) rotational constant $B_{e,\text{eff}}$. 
    The shaded bar indicates the range of $\Delta r_\text{{eq}}$ values consistent with our observations of the weak $v=3$ diagonal band.
        }
        \label{fig:Morse}
        \end{figure}

To understand the observed, unexpected intensity distribution of the rotational lines in the d$^3\Pi, v=3$ -- a$^3\Pi, v=3$ band as shown in Figure~\ref{fig:v3_full}, we investigate the effect of a perturbation of the d$^3\Pi$ state and the e$^3\Delta$ state. An interaction between these electronic states is the most obvious one to consider as the d$^3\Pi,v$ levels are only some 750\,cm$^{-1}$ below the e$^3\Delta, v$ levels and 160\,cm$^{-1}$ above the e$^3\Delta, v-1$ levels. In the following, we show that this perturbation can explain our observations. Moreover, a deperturbation analysis leads to a slight correction to the internuclear potentials of the d$^3\Pi$ and e$^3\Delta$ states. In Figure~\ref{fig:Morse}a the Morse potentials of both of these states as well as of the f$^3\Sigma^+$ state are shown, and the vibrational levels that have been characterized are drawn in. These Morse potentials are derived from the parameters $\omega_e$, $\omega_ex_e$ and $r_{\text{eq}}$ as given in Table~\ref{table:vibconsts}, with the value of $D_e$ taken as $\omega_e^2/(4\omega_e x_e)$. 

The perturbation between the e$^3\Delta$ and d$^3\Pi$ state can be due to spin-orbit interaction or spin-rotation interaction. The matrix elements for the perturbation between a $^3\Pi$ state and a $^3\Delta$ state, both described in Hund's case~(a) bases, depend on the $\Omega$-manifolds of the electronic states and the only non-zero ones are given by:\cite{1969Kovacsp}
\begin{align}
\label{eq:matrixelements}
M_{\Pi_0\Delta_1}&= 2 \eta \sqrt{J(J+1)}\\
M_{\Pi_1\Delta_2}&=2 \eta   \sqrt{(J-1)(J+2)}\\
M_{\Pi_2\Delta_3}&=2 \eta  \sqrt{(J-2)(J+3)}\\
M_{\Pi_1\Delta_1}=M_{\Pi_2\Delta_2}&=\sqrt{2}  (\xi/2+2\eta)
\label{eq:matrixelements2}
\end{align}
with the spin-orbit interaction parameter $\xi$ and the spin-rotation interaction parameter $\eta$. The interaction between specific vibrational levels $v_{\text{d}}$ and $v_{\text{e}}$ is given by the product of the vibrational overlap integral ($\mathcal{S}_{\text{de}}(v_{\text{d}}, v_{\text{e}})$) of those vibrational levels, i.e. by the square root of the Franck-Condon factor, with an overall spin-orbit ($A_{\text{so}}$) or spin-rotation ($E_{\text{sr}}$) interaction parameter between the electronic d$^3\Pi$ and e$^3\Delta$ states, that is
\begin{align}
\xi(v_{\text{d}}, v_{\text{e}}) &= A_{\text{so}} \, \mathcal{S}_{\text{de}}(v_{\text{d}}, v_{\text{e}}) \\
\eta(v_{\text{d}}, v_{\text{e}}) &= E_{\text{sr}} \, \mathcal{S}_{\text{de}}(v_{\text{d}}, v_{\text{e}}).
\end{align}

The Franck-Condon matrix between the e$^3\Delta$ state and the d$^3\Pi$ state is highly diagonal, i.e. $\mathcal{S}_{\text{de}}^2(v_{\text{d}}, v_{\text{e}})$ is close to one when $v_{\text{d}}$ = $v_{\text{e}}$ and rapidly decreases with increasing $|v_{\text{d}} - v_{\text{e}}|$. From the $B_{e,\text{eff}}$ values given in Table~\ref{table:vibconsts}, the internuclear equilibrium distance for the d$^3\Pi$ state would be determined as 1.5967(5)\,\AA, only slightly shorter than the equilibrium distance of 1.5994(7)\,\AA~for the e$^3\Delta$ state. In particular, the value for the Franck-Condon factor $\mathcal{S}_{\text{de}}^2(3,2)$ between the $v_{\text{d}}$ = 3 and $v_{\text{e}}$ = 2 levels would be expected to be only 0.009 in that case.

\subsubsection{Effect on the line intensities}

The transition intensity (square of the transition dipole moment) of the e$^3\Delta$ -- a$^3\Pi$ transition has been calculated to be about 20 times larger than that of the d$^3\Pi$ -- a$^3\Pi$ transition.\cite{Langhoff1988} The Franck-Condon factor for the e$^3\Delta, v=2$ -- a$^3\Pi, v=3$ band is about 0.30 (see Table~\ref{tab:fc}), and this band is the likely candidate that the intensity of the d$^3\Pi, v=3$ -- a$^3\Pi, v=3$ band is borrowed from. As indicated in Section~\ref{sec:res-e}, the strongest lines in the d$^3\Pi, v=3$ -- a$^3\Pi, v=3$ band are observed to have about $(2.5 \pm 0.5) \times 10^{-4}$ of the intensity of those in the e$^3\Delta, v=2$ -- a$^3\Pi, v=3$ band and lines with the expected intensity pattern for an allowed d$^3\Pi$ -- a$^3\Pi$ band are not observed at all. Given the signal-to-noise ratio in the spectra shown in Figure~\ref{fig:v3_full} of about 25, this implies that the transition intensity of the $v=3$ -- $v=3$ band of the d$^3\Pi$ -- a$^3\Pi$ transition is at least a factor 10$^5$ smaller than that of the e$^3\Delta, v=2$ -- a$^3\Pi, v=3$ band. When the transition dipole moment of the d$^3\Pi$ -- a$^3\Pi$ transition would be constant, i.e. independent of the internuclear distance, this would mean that the Franck-Condon factor of this diagonal band is less than $6 \times 10^{-5}$. As the electronic potentials of the e$^3\Delta$ and d$^3\Pi$ states are rather similar, the Franck-Condon factor for the diagonal e$^3\Delta, v=3$ -- a$^3\Pi, v=3$ band will also be small. Given that the e$^3\Delta, v=3$ state is predissociated, its vibrational wavefunction will be distorted. We therefore exclude the e$^3\Delta, v=3$ -- a$^3\Pi, v=3$ band when considering the intensity borrowing. 

In Figure~\ref{fig:Morse}b, the Franck-Condon factors for several of the diagonal bands of the d$^3\Pi$ -- a$^3\Pi$ transition are shown as a function of the equilibrium internuclear distance of the d$^3\Pi$ state. The zero on the horizontal axis corresponds to an equilibrium internuclear distance for the d$^3\Pi$ state of 1.5967\,\AA, which is the value extracted from the equilibrium rotational constant given in Table~\ref{table:vibconsts}. The internuclear equilibrium distance for the a$^3\Pi$ state is accurately known\cite{Walter2022} as $r_\text{{eq,a}}$ = 1.64708\,\AA~and is kept fixed at this value. When the equilibrium internuclear distance of the d$^3\Pi$ state is 1.5967\,\AA, the Franck-Condon factors for the $3-3$ and $4-4$ bands are seen to be very similar. Experimentally, both the $2-2$ and the $4-4$ bands are known to have transition intensities that are orders of magnitude larger than that of the $3-3$ band. Even though the dependence of the transition dipole moment on the internuclear distance can influence this somewhat, this indicates that the real equilibrium internuclear distance of the d$^3\Pi$ state is about 0.0027\,\AA~smaller than 1.5967\,\AA. 

To quantitatively model the relative line intensities observed in the spectra of the d$^3\Pi, v=3$ -- a$^3\Pi, v=3$ band, we diagonalize $6\times6$ matrices, set up on a Hund's case~(a) basis and including the three $\Omega$-manifolds of the two interacting states, for each $J$ value. As only transitions to a $^3\Delta_2$ manifold have non-zero H\"onl-London factors from a $^3\Pi_1$ manifold,\cite{Jenkins1934} we calculate the fraction of e$^3\Delta_2, v=2$ character in the wavefunction of the d$^3\Pi_{\Omega}, v=3$ state, denoted as $c_{\Delta_2}(\text{d}^3\Pi_{\Omega},J)$. To determine the $\xi(3,2)/\eta(3,2) = A_{\text{so}}/E_{\text{sr}}$ ratio that reproduces the experimentally observed relative line intensities the best, we define the intensity ratio $\mathcal{I}_{2\Omega}(J)$ as:
\begin{equation}
  \mathcal{I}_{2\Omega}(J)=
 \left(\frac{c_{\Delta_2} (\text{d}^3\Pi_2,J)}{c_{\Delta_2} (\text{d}^3\Pi_{\Omega},J)} \right)^2.
  \label{eq:ratioomega}
\end{equation}

\begin{figure}[]
        \centering
        \includegraphics[width = 480pt]{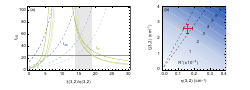}
        \caption{(a) $\mathcal{I}_{21}(J)$ (solid green lines) and $\mathcal{I}_{20}(J)$ (dashed blue lines) for $J=2-5$ (shown with decreasing color saturation), see Eq.~\ref{eq:ratioomega}.
    Values within the shaded area ($14<\xi(3,2)/\eta(3,2)<19$)
    agree with the experimental observation ($\mathcal{I}_{20},\mathcal{I}_{21}>25$).
    (b) Ratio of the intensities of the d$^3\Pi_2,v=3,J=2 \leftarrow \text{a}^3\Pi,v=3,J=1$ and e$^3\Delta_2,v=2,J=2 \leftarrow \text{a}^3\Pi,v=3,J=1$ lines
    that is experimentally found to be $(2.5 \pm 0.5) \times 10^{-4}$.
    The area between the dashed lines corresponds to the shaded area in (a).
     Both criteria are fulfilled for $\xi(3,2)=(2.6 \pm 0.3)$\,cm$^{-1}$ and $\eta(3,2)=(0.16 \pm 0.03)$\,cm$^{-1}$, indicated by the red point.}
        \label{fig:Pert}
        \end{figure}

This gives the ratio of the intensities of the lines that start from a certain rotational level in the a$^3\Pi_1, v=3$ state and that reach levels with the same value of $J$ in the d$^3\Pi_2, v=3$ and in the $\Omega=0, 1$ manifolds of the d$^3\Pi_{\Omega}, v=3$ state. A plot of $\mathcal{I}_{21}(J)$ and $\mathcal{I}_{20}(J)$ for $J=2-5$ as a function of the ratio $\xi(3,2)/\eta(3,2)$ is shown in Figure~\ref{fig:Pert}a. These curves are independent of the actual values of $\xi(3,2)$ and $\eta(3,2)$ when $\xi(3,2)$ $\leq$ 5 cm$^{-1}$ and $\eta(3,2)$ $\leq$ 0.3 cm$^{-1}$. Experimentally, lines to the $\Omega=2$ manifold are known to basically be the only ones observed, and the value of $\mathcal{I}_{21}(J)$ and $\mathcal{I}_{20}(J)$ thus has to be larger than about 25 for $J=2-5$. For this to hold, it is seen that the ratio of $\xi(3,2)/\eta(3,2)$ has to be between 14 and 19. The signs of $\xi(3,2)$ and $\eta(3,2)$ have to be the same; for negative values of the $\xi(3,2)/\eta(3,2)$ ratio the conditions on $\mathcal{I}_{21}(J)$ and $\mathcal{I}_{20}(J)$ cannot be met. 

In Figure~\ref{fig:Pert}b, the calculated ratio of the intensities of two lines, $\mathcal{R'}$, ending up in the $J=2$ level of the d$^3\Pi_2, v=3$ state and in the $J=2$ level in the e$^3\Delta_2, v=2$ state are shown. Both lines start from the same $J=1$ level in the a$^3\Pi_1, v=3$ state. The intensity ratio is shown as a function of the values of $\xi(3,2)$ and $\eta(3,2)$. The two criteria, namely that the $\xi(3,2)/\eta(3,2)$ ratio is between 14 and 19 and that the intensity ratio is between $2 \times 10^{-4}$ and $3 \times 10^{-4}$, are fulfilled for $\xi(3,2)=(2.6 \pm 0.3)$\,cm$^{-1}$ and $\eta(3,2)=(0.16 \pm 0.03)$\,cm$^{-1}$. 
The simulated spectrum of Figure~\ref{fig:v3_full}b is calculated using again Eq.~(\ref{eq:Idelta-pi}).
It is seen that in this way, the observed relative line intensities are well reproduced. 

\subsubsection{Effect on the line positions}
When the strength of the interaction between the e$^3\Delta, v=2$ and the d$^3\Pi, v=3$ state is on the order of magnitude of several cm$^{-1}$, one would expect to observe distortion of the regular rotational energy level structure. This is particularly so, because the strength of the interaction between levels with the same vibrational quantum number $v$ in the e$^3\Delta$ and d$^3\Pi$ state is larger with the factor $\mathcal{S}_{\text{de}}(v,v)/\mathcal{S}_{\text{de}}(3,2)$, whereas the energy separation of the interacting levels is only more by less than a factor five. The magnitude of the shift of the energy levels is given by the square of the interaction strength divided by the energy separation and the direction of the shift is such that the interacting levels repel each other. The downward shift of the ro-vibrational levels in the d$^3\Pi$ state due to spin-orbit interaction with levels with the same vibrational quantum number in the e$^3\Delta$ state, for instance, will thus approximately be $2.6^2/750 \times \mathcal{S}_{\text{de}}^2(v,v)/\mathcal{S}_{\text{de}}^2(3,2)$\,cm$^{-1}$. This seemed to be a contradiction at first, as no shift of the levels or distortion of the regular rotational structure of the levels have been observed whereas with the expected ratio of $\mathcal{S}_{\text{de}}^2(v,v)/\mathcal{S}_{\text{de}}^2(3,2)$ of about 100 these shifts should have been readily detectable. 

As the $J=0$ levels of the d$^3\Pi$ state are not influenced by the interaction with the e$^3\Delta$ state, any shift of the energy levels can be recognized most directly when transitions to this $J=0$ level are included in the spectra. Unfortunately, these transitions cannot be observed in the spectra recorded from the a$^3\Pi_1$ state as shown in Figure~\ref{fig:v0_full} and~\ref{fig:v3_full}. When recording transitions to the d$^3\Pi$ state via the b$^3\Sigma^+$ state, this lowest rotational level can be reached, as explicitly shown in Figure~\ref{fig:v3_full}. However, the spectral resolution is in that case limited to about 0.3\,cm$^{-1}$ due to the span of the hyperfine structure in the b$^3\Sigma^+$ state, preventing an accurate determination of a possible energy level shift.\cite{Doppelbauer2021} 

From the matrix elements for the perturbation given in Eq.~(\ref{eq:matrixelements}--\ref{eq:matrixelements2}), it is seen that for pure Hund's case~(a) states, the effect of spin-orbit interaction is $J$-independent and absent for $\Omega=0$ levels. The shift of the levels in the d$^3\Pi$ state due to the spin-orbit interaction with the e$^3\Delta$ state will therefore mainly be absorbed in the values of $A_v$ and $\lambda_v$ of the d$^3\Pi$ state. Both the spin-orbit and the spin-rotation interaction will also influence the term-values $E_{v}$, but as this effect is expected to be very similar for the lowest four vibrational levels of the d$^3\Pi$ state, this will go largely unnoticed. The effect of spin-rotation interaction will largely be absorbed in the value of the rotational constant. The effective rotational constants for the d$^3\Pi, v=0-3$ levels that are extracted from the spectra will be slightly less than the deperturbed rotational constants, i.e. the values expected purely due to end-over-end rotation. The rotational constant $B_0$ listed in Table~\ref{table:dstaterotconst}, for instance, will be reduced by about $(2\eta(0,0))^2/(750 \text{ cm}^{-1}$) relative to its deperturbed value. 
The deperturbed equilibrium rotational constant for the d$^3\Pi$ state, $B_{e,\text{depert}}$, will be larger by about this amount than the $B_\text{e,eff}$ value (see Table~\ref{table:vibconsts}). 
For the e$^3\Delta, v=0$ state, the shift of the rotational levels will be equal in magnitude but in opposite direction and
the $B_{e,\text{depert}}$ value of the e$^3\Delta$ state is less by $(2\eta(0,0))^2/(750 \text{ cm}^{-1}$) than the $B_{e,\text{eff}}$ value. As the equilibrium internuclear distances need to be calculated from the $B_{e,\text{depert}}$ values, $r_\text{{eq}}$(d$^3\Pi$) will be slightly shorter than the value of 1.5967(5)\,\AA, while $r_\text{{eq}}$(e$^3\Delta$) will be the same amount larger than the value of 1.5994(7)\,\AA~extracted from the $B_{e,\text{eff}}$ value given in Table~\ref{table:vibconsts}. 

\subsubsection{Summary on the perturbation analysis}

From the observation that the d$^3\Pi, v=3$ -- a$^3\Pi, v=3$ band has near zero transition intensity and from the Franck-Condon factors shown in Figure~\ref{fig:Morse}b, it is concluded that the real equilibrium internuclear distance of the d$^3\Pi$ state is 0.0027\,\AA~ smaller than 1.5967\,\AA. The unexpected pattern of borrowed intensities is well explained by a perturbation of the d$^3\Pi$ state with the e$^3\Delta$ state. This, combined with the equal but opposite energy levels shift due to the perturbation, puts the real equilibrium internuclear distance of the e$^3\Delta$ state at a 0.0027\,\AA~larger value than 1.5994\,\AA. This has a significant effect on the off-diagonal Franck-Condon factors between the d$^3\Pi$ state and the e$^3\Delta$ state; as can be seen from Table~\ref{tbl:FCde}, the value for $\mathcal{S}_{\text{de}}^2(3,2)$ is found to be 0.043 instead of 0.009.

\begin{table} \centering
  \caption{Franck-Condon factors $\mathcal{S}_{\text{de}}^2$ between the
   d$^3\Pi$ and e$^3\Delta$ states, calculated using Morse potentials derived from the parameters $\omega_e$, $\omega_ex_e$ and $r_{\text{eq}}$ as given in Table~\ref{table:vibconsts}.
  }
  \label{tbl:FCde}
  \begin{tabular}{lllllll}
    \hline
 &   $v_\text{d}=0$ &     $v_\text{d}=1$ &   $v_\text{d}=2$ &   $v_\text{d}=3$ &   $v_\text{d}=4$ & $v_\text{d}=5$  \\
     \hline 
$v_\text{e}=0$  & 0.988 & 0.012 & 0.000 & 0.000 & 0.000 & 0.000 \\ 
$v_\text{e}=1$ & 0.011 & 0.963 & 0.026 & 0.000 & 0.000 & 0.000 \\ 
$v_\text{e}=2$ &0.000 & 0.024 & 0.932 & \bf{0.043} & 0.000 & 0.000 \\ 
$v_\text{e}=3^\text{a}$& 0.000 & 0.001 & 0.040 & 0.895 & 0.064 & 0.000 \\ 

\hline
  \end{tabular}\\
  \textsuperscript{\emph{a}} Predissociated
\end{table}

From the analysis of the line intensities, the values of $\xi(3,2)=(2.6 \pm 0.3)$\,cm$^{-1}$ and $\eta(3,2)=(0.16 \pm 0.03)$\,cm$^{-1}$ are found. Together with the value for $\mathcal{S}_{\text{de}}(3,2)$ this results in values for the overall spin-orbit and spin-rotation interaction parameters between the electronic d$^3\Pi$ and e$^3\Delta$ states of $A_{\text{so}} = (12.5 \pm 1.5)$\,cm$^{-1}$ and $E_{\text{sr}} = (0.77 \pm 0.15 )$\,cm$^{-1}$.

It is seen from Table~\ref{table:vibconsts} that the deperturbed rotational constants for the d$^3\Pi$ state and the e$^3\Delta$ state differ from the effective rotational constants by 0.0020 cm$^{-1}$. It has been argued above, that this difference is approximately given by $(2\eta(0,0))^2/(750\text{ cm}^{-1})$.  As the value of $\mathcal{S}_{\text{de}}(0,0)$ is very close to one, this means that we find in this way that $E_{\text{sr}}$ is about 0.61\,cm$^{-1}$, yielding a fully consistent picture.

\section{Conclusions} 

In this study we have spectroscopically characterized the lowest seven vibrational levels of the newly found d$^3\Pi$ state of AlF and we have unambiguously identified the electronic state just above that as a $^3\Delta$ state. The lowest three vibrational levels of this e$^3\Delta$ state have been characterised, whereas only some broad remnants of the $v=3$ level could be detected; this vibrational level apparently predissociates on a timescale of a few picoseconds. We have characterised the $v=1$ and $v=2$ levels of a yet higher lying $^3\Sigma^+$ state, whose $v=0$ level had already been reported upon earlier, and this state should be referred to as the f$^3\Sigma^+$ state from now on. The radiative lifetimes of the d$^3\Pi$ and f$^3\Sigma^+$ states have been found as $(40 \pm 4)$\,ns and as $(30 \pm 6)$\,ns, whereas the lifetime of the e$^3\Delta$ state was found to be shorter than 10\,ns and thereby too short to be measured exactly with the present setup. The experimental data on these three electronic states of AlF are summarised in Table~\ref{table:vibconsts}. By ionization from the d$^3\Pi$ state, the ionization potential of AlF has been accurately determined as $(9.73177 \pm 0.00012$)\,eV. 

A most interesting observation that has been made is that the transition intensity of the d$^3\Pi, v=3$ -- a$^3\Pi, v=3$ band is close to zero. This peculiarity of a missing diagonal band might have contributed to the d$^3\Pi$ state not having been reported upon earlier. The observation of near zero transition intensity makes an accurate determination of the difference in equilibrium internuclear distance of the d$^3\Pi$ state and the a$^3\Pi$ state possible, namely by evaluating where the Franck-Condon factor for this band goes to zero. It is unlikely that
any contribution from the internuclear distance dependence of the transition dipole moment
would significantly contribute to the difference in equilibrium bond distance extracted from the
intensity analysis presented here.
Due to the negligible intensity of the $v=3$ -- $v=3$ band of the d$^3\Pi$ -- a$^3\Pi$ transition, we could recognize the very weak intensity that is borrowed from the nearby and strong e$^3\Delta, v=2$ -- a$^3\Pi, v=3$ band. This in turn enabled a detailed analysis of the spin-orbit and spin-rotation interaction between the d$^3\Pi$ state and the e$^3\Delta$ state, from which a spin-orbit interaction parameter $A_{\text{so}}$ of about 12.5\,cm$^{-1}$ and a spin-rotation interaction parameter $E_{\text{sr}}$ of about 0.8\,cm$^{-1}$ have been determined. These parameters, which might have error bars of up to 20\,\%, reproduce the borrowed rotational line intensities very well. The shift of the energy levels that is caused by the spin-rotation interaction is mainly absorbed into slightly perturbed values of the equilibrium rotational constants in the d$^3\Pi$ and e$^3\Delta$ states. After correction for the spin-rotation interaction, the real equilibrium internuclear distances can be extracted from the deperturbed rotational constants, and these agree with those expected from the Franck-Condon analysis. 

In their overview paper, Barrow, Klopp and Malmberg report on the observation of the displacement of a line in the spectrogram of the 
f$^3\Sigma^+,v=0$ -- c$^3\Sigma^+,v=0$ band.\cite{Barrow1974a} 
It appears that the $N=41$ level of the f$^3\Sigma^+, v=0$ state is shifted to lower energy by about 0.35(5)\,cm$^{-1}$, whereas the neighboring levels are hardly affected. From the parameters given in Table~\ref{table:estaterotconst}, we find that the $N=41$ level (in Hund's case~(b) notation) of the  e$^3\Delta, v=2$ state is near degenerate with the $N=41$ level of the f$^3\Sigma^+, v=0$ state. As the e$^3\Delta, v=2$ state has some $^3\Pi$ character mixed in, it will be the interaction of these near degenerate $N=41$ levels that causes this observed line displacement.

\section{Supporting Information} 

\begin{longtable}{@{\extracolsep{\fill}}lrccccc@{}}
\caption{\label{tab:linelist-d} Observed energies $E_{\text{exp}}$ (in cm$^{-1}$) of the d$^3\Pi_{\Omega_{\text{d}}},v_{\text{d}},J_{\text{d}}$ levels, reached via the $\text{a}^3\Pi_{\Omega_{\text{a}}},v_{\text{a}}$ states.
 All quantum numbers are given in the basis of Hund's case~(a).
}\\
\hline
\multicolumn{1}{l}{$E_{\text{exp}}$} &
 \multicolumn{1}{l}{$E_{\text{exp}}-E_{\text{calc}}$} &
      \multicolumn{1}{c}{$v_{\text{d}}$}&
           \multicolumn{1}{c}{$\Omega_{\text{d}}$}& 
             \multicolumn{1}{c}{$J_{\text{d}}$}&
      \multicolumn{1}{c}{$v_{\text{a}}$}&
           \multicolumn{1}{c}{$\Omega_{\text{a}}$} \\
\endfirsthead
\multicolumn{7}{c}%
        {{Table \thetable\ Continued from previous column.}} \\
\hline
\multicolumn{1}{l}{$E_{\text{exp}}$} &
 \multicolumn{1}{l}{$E_{\text{exp}}-E_{\text{calc}}$} &
      \multicolumn{1}{c}{$v_{\text{d}}$}&
           \multicolumn{1}{c}{$\Omega_{\text{d}}$}& 
             \multicolumn{1}{c}{$J_{\text{d}}$}&
      \multicolumn{1}{c}{$v_{\text{a}}$}&
           \multicolumn{1}{c}{$\Omega_{\text{a}}$} \\[0.1cm]
\hline
\endhead
\hline
62901.398 & 0.003  &  0 & 0 & 1  &  0 & 1 \\
62903.277 & $-$0.035  &  0 & 0 & 2  &  0 & 1 \\
62906.219 & $-$0.012  &  0 & 0 & 3  &  0 & 1 \\
62907.621 & $-$0.024  &  0 & 1 & 1  &  0 & 1 \\
62909.695 & 0.002  &  0 & 1 & 2  &  0 & 1 \\
62910.223 & 0.031  &  0 & 0 & 4  &  0 & 1 \\
62912.977 & 0.009  &  0 & 1 & 3  &  0 & 1 \\
62913.891 & 0.026  &  0 & 2 & 2  &  0 & 1 \\
62915.199 & $-$0.029  &  0 & 0 & 5  &  0 & 1 \\
62917.555 & 0.065  &  0 & 1 & 4  &  0 & 1 \\
62918.285 & $-$0.025  &  0 & 2 & 3  &  0 & 1 \\
62921.418 & 0.050  &  0 & 0 & 6  &  0 & 1 \\
62923.250 & 0.004  &  0 & 1 & 5  &  0 & 1 \\
62923.969 & $-$0.040  &  0 & 2 & 4  &  0 & 1 \\
62930.191 & $-$0.034  &  0 & 1 & 6  &  0 & 1 \\
62930.973 & 0.028  &  0 & 2 & 5  &  0 & 1 \\
62939.074 & $-$0.018  &  0 & 2 & 6  &  0 & 1 \\
63835.234 & $-$0.019  &  1 & 0 & 1  &  1 & 1 \\
63837.180 & 0.027  &  1 & 0 & 2  &  1 & 1 \\
63840.051 & 0.007  &  1 & 0 & 3  &  1 & 1 \\
63841.395 & $-$0.026  &  1 & 1 & 1  &  1 & 1 \\
63843.527 & 0.049  &  1 & 1 & 2  &  1 & 1 \\
63843.938 & $-$0.025  &  1 & 0 & 4  &  1 & 1 \\
63846.770 & 0.029  &  1 & 1 & 3  &  1 & 1 \\
63847.742 & 0.026  &  1 & 2 & 2  &  1 & 1 \\
63848.988 & 0.033  &  1 & 0 & 5  &  1 & 1 \\
63851.164 & $-$0.069  &  1 & 1 & 4  &  1 & 1 \\
63852.043 & $-$0.055  &  1 & 2 & 3  &  1 & 1 \\
63855.008 & $-$0.033  &  1 & 0 & 6  &  1 & 1 \\
63856.949 & 0.006  &  1 & 1 & 5  &  1 & 1 \\
63857.746 & 0.012  &  1 & 2 & 4  &  1 & 1 \\
63863.895 & 0.032  &  1 & 1 & 6  &  1 & 1 \\
63864.598 & 0.006  &  1 & 2 & 5  &  1 & 1 \\
63872.656 & $-$0.001  &  1 & 2 & 6  &  1 & 1 \\
64758.895 & 0.011  &  2 & 0 & 1  &  2 & 1 \\
64760.770 & 0.000  &  2 & 0 & 2  &  2 & 1 \\
64763.648 & 0.008  &  2 & 0 & 3  &  2 & 1 \\
64765.117 & 0.046  &  2 & 1 & 1  &  2 & 1 \\
64767.078 & $-$0.044  &  2 & 1 & 2  &  2 & 1 \\
64767.520 & $-$0.015  &  2 & 0 & 4  &  2 & 1 \\
64770.340 & $-$0.023  &  2 & 1 & 3  &  2 & 1 \\
64771.379 & $-$0.030  &  2 & 2 & 2  &  2 & 1 \\
64772.465 & $-$0.021  &  2 & 0 & 5  &  2 & 1 \\
64774.809 & $-$0.008  &  2 & 1 & 4  &  2 & 1 \\
64775.781 & 0.046  &  2 & 2 & 3  &  2 & 1 \\
64778.512 & $-$0.010  &  2 & 0 & 6  &  2 & 1 \\
64780.520 & 0.043  &  2 & 1 & 5  &  2 & 1 \\
64781.332 & 0.025  &  2 & 2 & 4  &  2 & 1 \\
64787.371 & 0.038  &  2 & 1 & 6  &  2 & 1 \\
64788.027 & $-$0.070  &  2 & 2 & 5  &  2 & 1 \\
64796.086 & 0.003  &  2 & 2 & 6  &  2 & 1 \\
65672.359 & 0.005  &  3 & 0 & 1  &  2 & 1 \\
65674.188 & $-$0.037  &  3 & 0 & 2  &  2 & 1 \\
65677.047 & $-$0.034  &  3 & 0 & 3  &  2 & 1 \\
65678.438 & $-$0.051  &  3 & 1 & 1  &  2 & 1 \\
65680.609 & 0.064  &  3 & 1 & 2  &  2 & 1 \\
65681.000 & 0.053  &  3 & 0 & 4  &  2 & 1 \\
65683.789 & 0.016  &  3 & 1 & 3  &  2 & 1 \\
65684.898 & 0.022  &  3 & 2 & 2  &  2 & 1 \\
65685.867 & 0.004  &  3 & 0 & 5  &  2 & 1 \\
65688.180 & $-$0.024  &  3 & 1 & 4  &  2 & 1 \\
65689.148 & $-$0.015  &  3 & 2 & 3  &  2 & 1 \\
65691.891 & 0.036  &  3 & 0 & 6  &  2 & 1 \\
65693.820 & $-$0.013  &  3 & 1 & 5  &  2 & 1 \\
65694.695 & 0.010  &  3 & 2 & 4  &  2 & 1 \\
65700.625 & $-$0.020  &  3 & 1 & 6  &  2 & 1 \\
65701.398 & $-$0.024  &  3 & 2 & 5  &  2 & 1 \\
65709.352 & 0.008  &  3 & 2 & 6  &  2 & 1 \\
66575.156 & 0.013  &  4 & 0 & 1  &  4 & 1 \\
66576.984 & $-$0.020  &  4 & 0 & 2  &  4 & 1 \\
66579.844 & 0.002  &  4 & 0 & 3  &  4 & 1 \\
66581.328 & 0.009  &  4 & 1 & 1  &  4 & 1 \\
66583.367 & 0.009  &  4 & 1 & 2  &  4 & 1 \\
66583.672 & $-$0.012  &  4 & 0 & 4  &  4 & 1 \\
66586.531 & $-$0.033  &  4 & 1 & 3  &  4 & 1 \\
66587.703 & 0.005  &  4 & 2 & 2  &  4 & 1 \\
66588.570 & 0.002  &  4 & 0 & 5  &  4 & 1 \\
66590.969 & 0.008  &  4 & 1 & 4  &  4 & 1 \\
66591.938 & $-$0.005  &  4 & 2 & 3  &  4 & 1 \\
66594.539 & 0.016  &  4 & 0 & 6  &  4 & 1 \\
66596.531 & $-$0.006  &  4 & 1 & 5  &  4 & 1 \\
66597.445 & 0.023  &  4 & 2 & 4  &  4 & 1 \\
66603.312 & 0.021  &  4 & 1 & 6  &  4 & 1 \\
66604.078 & $-$0.018  &  4 & 2 & 5  &  4 & 1 \\
66611.938 & $-$0.015  &  4 & 2 & 6  &  4 & 1 \\
67469.461 & 0.049  &  5 & 0 & 2  &  5 & 1 \\
67472.180 & $-$0.055  &  5 & 0 & 3  &  5 & 1 \\
67473.836 & $-$0.009  &  5 & 1 & 1  &  5 & 1 \\
67475.844 & $-$0.006  &  5 & 1 & 2  &  5 & 1 \\
67476.117 & 0.064  &  5 & 0 & 4  &  5 & 1 \\
67479.008 & 0.006  &  5 & 1 & 3  &  5 & 1 \\
67480.062 & $-$0.049  &  5 & 2 & 2  &  5 & 1 \\
67480.898 & $-$0.008  &  5 & 0 & 5  &  5 & 1 \\
67483.344 & 0.001  &  5 & 1 & 4  &  5 & 1 \\
67484.344 & 0.023  &  5 & 2 & 3  &  5 & 1 \\
67486.734 & $-$0.089  &  5 & 0 & 6  &  5 & 1 \\
67488.953 & 0.096  &  5 & 1 & 5  &  5 & 1 \\
67489.727 & $-$0.026  &  5 & 2 & 4  &  5 & 1 \\
67495.508 & $-$0.034  &  5 & 1 & 6  &  5 & 1 \\
67496.398 & 0.025  &  5 & 2 & 5  &  5 & 1 \\
67504.172 & 0.013  &  5 & 2 & 6  &  5 & 1 \\
68351.08 &   $-$0.03  & 6  &  0     &   2       & 6 &   1 \\
68353.80  &  $-$0.11  & 6   & 0     &   3       & 6 &   1 \\
68355.57  &   0.05  & 6  &  1    &    1        & 6 &   1 \\
68357.47  &  $-$0.06  & 6  &  1    &    2         & 6 &   1 \\
68357.75  &   0.06   &6   & 0    &    4        & 6 &   1 \\
68360.59  &  $-$0.07  & 6  &  1   &     3        & 6 &   1 \\
68361.97  &   0.08  & 6   & 2    &    2        & 6 &   1 \\
68362.43 &   $-$0.07  & 6  &  0    &    5        & 6 &   1 \\
68365.01  &   0.03  & 6   & 1    &    4        & 6 &   1 \\
68365.98 &   $-$0.06  & 6 &   2     &   3        & 6 &   1 \\
68368.57   &  0.22  & 6  &  0    &    6        & 6 &   1 \\
68370.52  &   0.08  & 6  &  1    &    5        & 6 &   1 \\
68371.41  &   0.02  & 6  &  2    &    4        & 6 &   1 \\
68376.98   & $-$0.08 &  6  &  1    &    6        & 6 &   1 \\
68378.01  &   0.08 &  6  &  2    &    5        & 6 &   1 \\
68385.50  &  $-$0.13 &  6   & 2    &    6        & 6 &   1 \\
\hline
\end{longtable}

\begin{longtable}{@{\extracolsep{\fill}}lrccccc@{}}
\caption{\label{tab:linelist-e} 
Observed energies $E_{\text{exp}}$ (in cm$^{-1}$) of the e$^3\Delta_{\Omega_{\text{e}}},v_{\text{e}},J_{\text{e}}$ levels, reached via the $\text{a}^3\Pi_{\Omega_{\text{a}}},v_{\text{a}}$ states.
 All quantum numbers are given in the basis of Hund's case~(a).
}\\
\hline
\multicolumn{1}{l}{$E_{\text{exp}}$} &
 \multicolumn{1}{l}{$E_{\text{exp}}-E_{\text{calc}}$} &
      \multicolumn{1}{c}{$v_{\text{e}}$}&
           \multicolumn{1}{c}{$\Omega_{\text{e}}$}& 
             \multicolumn{1}{c}{$J_{\text{e}}$}&
      \multicolumn{1}{c}{$v_{\text{a}}$}&
           \multicolumn{1}{c}{$\Omega_{\text{a}}$} \\
\endfirsthead
\multicolumn{7}{c}%
        {{Table \thetable\ 
       Continued from previous column.}} \\
\hline
\multicolumn{1}{l}{$E_{\text{exp}}$} &
 \multicolumn{1}{l}{$E_{\text{exp}}-E_{\text{calc}}$} &
      \multicolumn{1}{c}{$v_{\text{e}}$}&
           \multicolumn{1}{c}{$\Omega_{\text{e}}$}& 
             \multicolumn{1}{c}{$J_{\text{e}}$}&
      \multicolumn{1}{c}{$v_{\text{a}}$}&
           \multicolumn{1}{c}{$\Omega_{\text{a}}$} \\[0.1cm]
\hline
\endhead
\hline
63674.125 & 0.020 & 0 & 3 & 3 & 1 & 1 \\
63674.945 & 0.018 & 0 & 2 & 2 & 1 & 1 \\
63675.473 & $-$0.031 & 0 & 1 & 1 & 1 & 0 \\
63677.766 & $-$0.003 & 0 & 3 & 4 & 1 & 1 \\
63678.348 & $-$0.024 & 0 & 2 & 3 & 1 & 1 \\
63678.836 & 0.028 & 0 & 1 & 2 & 1 & 1 \\
63682.566 & 0.001 & 0 & 3 & 5 & 1 & 1 \\
63683.023 & $-$0.010 & 0 & 2 & 4 & 1 & 1 \\
63683.387 & 0.001 & 0 & 1 & 3 & 1 & 1 \\
63688.512 & $-$0.004 & 0 & 3 & 6 & 1 & 1 \\
63688.883 & $-$0.010 & 0 & 2 & 5 & 1 & 1 \\
63689.215 & 0.022 & 0 & 1 & 4 & 1 & 1 \\
63695.891 & $-$0.049 & 0 & 2 & 6 & 1 & 1 \\
63696.223 & 0.024 & 0 & 1 & 5 & 1 & 1 \\
63704.418 & 0.016 & 0 & 1 & 6 & 1 & 1 \\
64602.156 & $-$0.031 & 1 & 3 & 3 & 2 & 1 \\
64603.141 & 0.014 & 1 & 2 & 2 & 2 & 1 \\
64603.793 & 0.001 & 1 & 1 & 1 & 2 & 0 \\
64605.887 & 0.048 & 1 & 3 & 4 & 2 & 1 \\
64606.500 & $-$0.033 & 1 & 2 & 3 & 2 & 1 \\
64607.031 & $-$0.008 & 1 & 1 & 2 & 2 & 1 \\
64610.598 & $-$0.013 & 1 & 3 & 5 & 2 & 1 \\
64611.137 & $-$0.016 & 1 & 2 & 4 & 2 & 1 \\
64611.582 & 0.021 & 1 & 1 & 3 & 2 & 1 \\
64616.547 & 0.029 & 1 & 3 & 6 & 2 & 1 \\
64616.953 & $-$0.006 & 1 & 2 & 5 & 2 & 1 \\
64617.328 & 0.020 & 1 & 1 & 4 & 2 & 1 \\
64623.922 & $-$0.026 & 1 & 2 & 6 & 2 & 1 \\
64624.254 & 0.005 & 1 & 1 & 5 & 2 & 1 \\
64632.371 & $-$0.004 & 1 & 1 & 6 & 2 & 1 \\
65516.965 & 0.010 & 2 & 3 & 3 & 3 & 1 \\
65518.188 & $-$0.026 & 2 & 2 & 2 & 3 & 1 \\
65519.098 & $-$0.026 & 2 & 1 & 1 & 3 & 0 \\
65520.684 & 0.053 & 2 & 3 & 4 & 3 & 1 \\
65521.527 & $-$0.052 & 2 & 2 & 3 & 3 & 1 \\
65522.277 & 0.004 & 2 & 1 & 2 & 3 & 1 \\
65525.406 & 0.007 & 2 & 3 & 5 & 3 & 1 \\
65526.199 & 0.048 & 2 & 2 & 4 & 3 & 1 \\
65526.734 & 0.017 & 2 & 1 & 3 & 3 & 1 \\
65531.242 & $-$0.046 & 2 & 3 & 6 & 3 & 1 \\
65531.875 & $-$0.030 & 2 & 2 & 5 & 3 & 1 \\
65532.465 & 0.079 & 2 & 1 & 4 & 3 & 1 \\
65538.766 & $-$0.062 & 2 & 2 & 6 & 3 & 1 \\
65539.250 & 0.004 & 2 & 1 & 5 & 3 & 1 \\
65547.320 & 0.020 & 2 & 1 & 6 & 3 & 1 \\
\hline
\end{longtable}

\begin{longtable}{@{\extracolsep{\fill}}lrccccc@{}}
\caption{\label{tab:linelist-f} Observed energies $E_{\text{exp}}$ (in cm$^{-1}$) of the f$^3\Sigma^+,v_{\text{f}},N_{\text{f}},p_{\text{f}}$ levels, reached via the $\text{a}^3\Pi_{\Omega_{\text{a}}},v_{\text{a}}$ states.
$E_{\text{exp}}$ is the absolute energy of the reached level.
The quantum numbers of the f$^3\Sigma^+$ state are given in the basis of Hund's case~(b),
the quantum numbers of the a$^3\Pi$ state in the basis of Hund's case~(a).
}\\
\hline
\multicolumn{1}{l}{$E_{\text{exp}}$} &
 \multicolumn{1}{l}{$E_{\text{exp}}-E_{\text{calc}}$} &
      \multicolumn{1}{c}{$v_{\text{f}}$}&
           \multicolumn{1}{c}{$N_{\text{f}}$}& 
             \multicolumn{1}{c}{$p_{\text{f}}$}&
      \multicolumn{1}{c}{$v_{\text{a}}$}&
           \multicolumn{1}{c}{$\Omega_{\text{a}}$} \\
\endfirsthead
\multicolumn{7}{c}%
        {{Table \thetable\ Continued from previous column.}} \\
\hline
\multicolumn{1}{l}{$E_{\text{exp}}$} &
 \multicolumn{1}{l}{$E_{\text{exp}}-E_{\text{calc}}$} &
      \multicolumn{1}{c}{$v_{\text{f}}$}&
           \multicolumn{1}{c}{$N_{\text{f}}$}& 
             \multicolumn{1}{c}{$p_{\text{f}}$}&
      \multicolumn{1}{c}{$v_{\text{a}}$}&
           \multicolumn{1}{c}{$\Omega_{\text{a}}$} \\[0.1cm]
\hline
\endhead
\hline
65488.176 & 0.002 & 0 & 0 &$+$  &  3 & 1 \\
65489.365 & 0.002 & 0 & 1 &$-$  &  3 & 1 \\
65491.717 & $-$0.023 & 0 & 2 &$+$  &  3 & 1 \\
65495.341 & 0.035 & 0 & 3 &$-$   &  3 & 1 \\
65500.039 & $-$0.022 & 0 & 4 &$+$  &  3 & 1 \\
65506.009 & 0.005 & 0 & 5 &$-$   &  3 & 1 \\
65513.136 & $-$0.000 & 0 & 6 &$+$  &  3 & 1 \\
66421.520 & 0.020 & 1 & 0 &$+$  &  4 & 1 \\
66422.674 & $-$0.011 & 1 & 1 &$-$   &  4 & 1 \\
66425.027 & $-$0.028 & 1 & 2 &$+$  &  4 & 1 \\
66428.579 & $-$0.033 & 1 & 3 &$-$   &  4 & 1 \\
66433.412 & 0.057 & 1 & 4 &$+$  &  4 & 1 \\
66439.310 & 0.028 & 1 & 5 &$-$   &  4 & 1 \\
66446.362 & $-$0.033 & 1 & 6 &$+$  &  4 & 1 \\
67343.200 & 0.042 & 2 & 0 &$+$  &  2 & 4 \\
67344.260 & $-$0.062 & 2 & 1 &$-$   &  4 & 1 \\
67346.580 & $-$0.070 & 2 & 2 &$+$  &  4 & 1 \\
67350.230 & 0.089 & 2 & 3 &$-$   &  4 & 1 \\
67354.770 & $-$0.026 & 2 & 4 &$+$  &  4 & 1 \\
67360.710 & 0.095 & 2 & 5 &$-$   &  4 & 1 \\
67367.530 & $-$0.068 & 2 & 6 &$+$  &  4 & 1 \\
\hline
\end{longtable}

\begin{longtable}{@{\extracolsep{\fill}}ccccrrrrrr@{}}
\caption{\label{tab:linelist-f} Wavefunction characters of the d$^3\Pi,v=0$, $v=3$ and  e$^3\Delta,v=2$ states.}\\
\hline
\multicolumn{1}{c}{state} &
 \multicolumn{1}{c}{$v$} &
      \multicolumn{1}{c}{$\Omega$}&
           \multicolumn{1}{c}{$J$}& 
             \multicolumn{1}{c}{$c_{\Pi_0}$}&
      \multicolumn{1}{c}{$c_{\Pi_1}$}&
      \multicolumn{1}{c}{$c_{\Pi_2}$}&
      \multicolumn{1}{c}{$c_{\Delta_1}$}&
      \multicolumn{1}{c}{$c_{\Delta_2}$}&
      \multicolumn{1}{c}{$c_{\Delta_3}$}
      \\
\endfirsthead
\multicolumn{10}{c}%
        {{Table \thetable\ Continued from previous column.}} \\
\hline
\multicolumn{1}{c}{state} &
 \multicolumn{1}{c}{$v$} &
      \multicolumn{1}{c}{$\Omega$}&
           \multicolumn{1}{c}{$J$}& 
             \multicolumn{1}{c}{$c_{\Pi_0}$}&
      \multicolumn{1}{c}{$c_{\Pi_1}$}&
      \multicolumn{1}{c}{$c_{\Pi_2}$}&
      \multicolumn{1}{c}{$c_{\Delta_1}$}&
      \multicolumn{1}{c}{$c_{\Delta_2}$}&
      \multicolumn{1}{c}{$c_{\Delta_3}$}\\[0.1cm]
\hline
\endhead
\hline
d$^3\Pi$ & 0 & 0 & 0 & 1.000 & 0.000 & 0.000 & $-$ & $-$ & $-$ \\
d$^3\Pi$ & 0 & 0 & 1 & 0.981 & $-$0.193 & 0.000 & $-$ & $-$ & $-$ \\
d$^3\Pi$ & 0 & 0 & 2 & 0.948 & $-$0.314 & 0.054 & $-$ & $-$ & $-$ \\
d$^3\Pi$ & 0 & 0 & 3 & 0.907 & $-$0.409 & 0.104 & $-$ & $-$ & $-$ \\
d$^3\Pi$ & 0 & 1 & 1 & 0.193 & 0.981 & 0.000 & $-$ & $-$ & $-$ \\
d$^3\Pi$ & 0 & 1 & 2 & 0.307 & 0.855 & $-$0.418 & $-$ & $-$ & $-$ \\
d$^3\Pi$ & 0 & 0 & 4 & 0.864 & $-$0.480 & 0.152 & $-$ & $-$ & $-$ \\
d$^3\Pi$ & 0 & 1 & 3 & 0.398 & 0.746 & $-$0.535 & $-$ & $-$ & $-$ \\
d$^3\Pi$ & 0 & 2 & 2 & 0.086 & 0.413 & 0.907 & $-$ & $-$ & $-$ \\
d$^3\Pi$ & 0 & 0 & 5 & $-$0.825 & 0.531 & $-$0.195 & $-$ & $-$ & $-$ \\
d$^3\Pi$ & 0 & 1 & 4 & 0.468 & 0.653 & $-$0.595 & $-$ & $-$ & $-$ \\
d$^3\Pi$ & 0 & 2 & 3 & 0.141 & 0.526 & 0.839 & $-$ & $-$ & $-$ \\
d$^3\Pi$ & 0 & 0 & 6 & 0.790 & $-$0.568 & 0.231 & $-$ & $-$ & $-$ \\
d$^3\Pi$ & 0 & 1 & 5 & $-$0.520 & $-$0.576 & 0.630 & $-$ & $-$ & $-$ \\
d$^3\Pi$ & 0 & 2 & 4 & 0.186 & 0.586 & 0.789 & $-$ & $-$ & $-$ \\
d$^3\Pi$ & 0 & 0 & 7 & 0.760 & $-$0.595 & 0.261 & $-$ & $-$ & $-$ \\
d$^3\Pi$ & 0 & 1 & 6 & 0.559 & 0.512 & $-$0.652 & $-$ & $-$ & $-$ \\
d$^3\Pi$ & 0 & 2 & 5 & 0.223 & 0.621 & 0.751 & $-$ & $-$ & $-$ \\
d$^3\Pi$ & 0 & 0 & 8 & 0.735 & $-$0.615 & 0.286 & $-$ & $-$ & $-$ \\
d$^3\Pi$ & 0 & 1 & 7 & $-$0.588 & $-$0.459 & 0.666 & $-$ & $-$ & $-$ \\
d$^3\Pi$ & 0 & 2 & 6 & 0.252 & 0.644 & 0.722 & $-$ & $-$ & $-$ \\
d$^3\Pi$ & 3 & 0 & 0 & 1.000 & 0.000 & 0.000 & 0.000 & 0.000 & 0.000 \\
d$^3\Pi$ & 3 & 0 & 1 & 0.982 & $-$0.188 & 0.000 & $-$0.001 & 0.000 & 0.000 \\
d$^3\Pi$ & 3 & 1 & 1 & 0.188 & 0.982 & 0.000 & $-$0.006 & 0.000 & 0.000 \\
d$^3\Pi$ & 3 & 0 & 2 & 0.950 & $-$0.307 & 0.050 & $-$0.001 & 0.000 & 0.000 \\
d$^3\Pi$ & 3 & 1 & 2 & 0.301 & 0.867 & $-$0.397 & $-$0.006 & 0.000 & 0.000 \\
d$^3\Pi$ & 3 & 2 & 2 & $-$0.078 & $-$0.393 & $-$0.916 & 0.003 & 0.006 & 0.000 \\
d$^3\Pi$ & 3 & 0 & 3 & 0.911 & $-$0.401 & 0.098 & $-$0.002 & 0.001 & 0.000 \\
d$^3\Pi$ & 3 & 1 & 3 & $-$0.390 & $-$0.762 & 0.517 & 0.006 & 0.000 & $-$0.002 \\
d$^3\Pi$ & 3 & 2 & 3 & 0.132 & 0.509 & 0.851 & $-$0.004 & $-$0.007 & $-$0.003 \\
d$^3\Pi$ & 3 & 0 & 4 & 0.870 & $-$0.471 & 0.144 & $-$0.002 & 0.002 & $-$0.001 \\
d$^3\Pi$ & 3 & 1 & 4 & $-$0.460 & $-$0.671 & 0.581 & 0.007 & 0.000 & $-$0.003 \\
d$^3\Pi$ & 3 & 2 & 4 & 0.177 & 0.572 & 0.801 & $-$0.005 & $-$0.008 & $-$0.004 \\
d$^3\Pi$ & 3 & 0 & 5 & 0.831 & $-$0.524 & 0.186 & $-$0.003 & 0.002 & $-$0.001 \\
d$^3\Pi$ & 3 & 1 & 5 & $-$0.513 & $-$0.594 & 0.619 & 0.007 & 0.000 & $-$0.004 \\
d$^3\Pi$ & 3 & 2 & 5 & 0.213 & 0.610 & 0.763 & $-$0.005 & $-$0.009 & $-$0.005 \\
d$^3\Pi$ & 3 & 0 & 6 & 0.797 & $-$0.562 & 0.222 & $-$0.003 & 0.003 & $-$0.002 \\
d$^3\Pi$ & 3 & 1 & 6 & $-$0.553 & $-$0.530 & 0.643 & 0.008 & 0.000 & $-$0.005 \\
d$^3\Pi$ & 3 & 2 & 6 & 0.244 & 0.635 & 0.733 & $-$0.006 & $-$0.009 & $-$0.005 \\
e$^3\Delta$ & 2 & 3 & 3 & $-$ & $-$ & $-$ & 0.126 & $-$0.388 & 0.913 \\
e$^3\Delta$ & 2 & 2 & 2 & $-$ & $-$ & $-$ & $-$0.453 & 0.892 & 0.000 \\
e$^3\Delta$ & 2 & 1 & 1 & $-$ & $-$ & $-$ & 1.000 & 0.000 & 0.000 \\
e$^3\Delta$ & 2 & 3 & 4 & $-$ & $-$ & $-$ & 0.195 & $-$0.500 & 0.844 \\
e$^3\Delta$ & 2 & 2 & 3 & $-$ & $-$ & $-$ & $-$0.565 & 0.728 & 0.388 \\
e$^3\Delta$ & 2 & 1 & 2 & $-$ & $-$ & $-$ & 0.892 & 0.453 & 0.000 \\
e$^3\Delta$ & 2 & 3 & 5 & $-$ & $-$ & $-$ & 0.244 & $-$0.560 & 0.791 \\
e$^3\Delta$ & 2 & 2 & 4 & $-$ & $-$ & $-$ & 0.618 & $-$0.606 & $-$0.501 \\
e$^3\Delta$ & 2 & 1 & 3 & $-$ & $-$ & $-$ & 0.815 & 0.565 & 0.127 \\
e$^3\Delta$ & 2 & 3 & 6 & $-$ & $-$ & $-$ & $-$0.281 & 0.597 & $-$0.751 \\
e$^3\Delta$ & 2 & 2 & 5 & $-$ & $-$ & $-$ & $-$0.646 & 0.514 & 0.564 \\
e$^3\Delta$ & 2 & 1 & 4 & $-$ & $-$ & $-$ & 0.762 & 0.619 & 0.191 \\
e$^3\Delta$ & 2 & 3 & 7 & $-$ & $-$ & $-$ & 0.309 & $-$0.621 & 0.720 \\
e$^3\Delta$ & 2 & 2 & 6 & $-$ & $-$ & $-$ & $-$0.663 & 0.445 & 0.602 \\
e$^3\Delta$ & 2 & 1 & 5 & $-$ & $-$ & $-$ & 0.723 & 0.649 & 0.237 \\
e$^3\Delta$ & 2 & 3 & 8 & $-$ & $-$ & $-$ & $-$0.331 & 0.638 & $-$0.696 \\
e$^3\Delta$ & 2 & 2 & 7 & $-$ & $-$ & $-$ & 0.674 & $-$0.391 & $-$0.626 \\
e$^3\Delta$ & 2 & 1 & 6 & $-$ & $-$ & $-$ & 0.694 & 0.667 & 0.271 \\
\hline
\end{longtable}

We acknowledge the technical support of Sebastian Kray and Klaus-Peter Vogelgesang. NW thanks Thorsten Brand for inspiring discussions.

\bibliography{acs-achemso.bib}




\end{document}